\documentclass[intlimits,twoside,a4paper]{article}

\usepackage[utf8]{inputenc}
\usepackage[eqsecnum]{cmpj3prep}
\usepackage[percent]{overpic}
\usepackage{graphicx}

\newcommand{\panel}[5]{%
  \begin{overpic}[width=#1]{#5}\put(#2,#3){\sffamily\bfseries\small(#4)}\end{overpic}}

\title[Competing-interaction fluid and a solvent near a hard wall]%
{Mixtures of a fluid with competing interactions and a simple solvent near a hard wall}
\author[I.~Kravtsiv, O.~Patsahan, T.~Patsahan]{I.~Kravtsiv\orcid{0000-0001-7344-9528}\refaddr{label1}\thanks{Corresponding author: \email{ivankr@icmp.lviv.ua}}, 
O.~Patsahan\orcid{ 0000-0002-5839-3893}\refaddr{label1},
T.~Patsahan\orcid{0000-0002-7870-2219}\refaddr{label1,label2}}
\addresses{
\addr{label1} Yukhnovskii Institute for Condensed Matter Physics of the National Academy of Sciences of Ukraine, 1~Svientsitskii Str., 79011, Lviv, Ukraine
\addr{label2} Institute of Applied Mathematics and Fundamental Sciences, Lviv Polytechnic National University, 12~S.~Bandera Str., 79013 Lviv, Ukraine
}


\begin{document}

\maketitle

\begin{abstract}
We study a binary mixture of soft particles near a planar surface. The main
component interacts through a competing short-range attraction and long-range
repulsion (SALR) potential, modelled by a sum of three Yukawa terms, while the
solvent and the cross interactions are described by two Yukawa terms. Using a
field-theoretical mean-field approach we derive the coupled equations for the
density profiles of both species, formulate the contact theorem for the
two-component system, and obtain closed-form profiles in the mesoscopically
homogeneous region by linearizing the theory and solving the associated
eigenvalue problem. Their amplitudes are fixed by exact boundary
conditions following from the continuity of the Yukawa potentials at the wall,
which supply as many relations as there are unknowns. Although these conditions
are implicit in the derivation of the contact theorem, they have not previously
been used to fix the amplitudes of the linearized profiles, and they bring those
profiles into quantitative agreement with the nonlinear solution. The equations are solved
numerically for a fluid confined by hard walls. The
theory predicts mixing of the two species far
from the surface at high temperatures, and their microphase separation
into alternating layers upon cooling or upon increasing the density. The mean-field profiles agree well with molecular dynamics data once the mean-field overestimate of the onset temperature of microsegregation is taken
into account, and the analytical ones reproduce them closely except in the
immediate vicinity of the wall.
\printkeywords
\pacs 61.20.Gy, 68.08.-p, 61.20.Ja, 05.20.Jj
\end{abstract}

\section{Introduction}
\label{sec:intro}

Biological and soft-matter systems are usually spatially inhomogeneous and
multicomponent, since they are frequently composed of a mixture of charged
macromolecules or nanoparticles immersed in a medium of much smaller solvent
molecules. Despite the substantial progress achieved in the study of
one-component fluids with competing interactions, mixtures of particles with
competing interactions started to attract attention only
recently~\cite{Sweatman2019,vonKonigslow_2013,FerreiroRangel2018,Tan2021,TanSweatman2024,Munao2022,munao:22:0,PatsahanLitniewskiCiach2021,PatsahanMeyraCiach2022,PatsahanMeyraCiach2024,Prestipino2023,Pini2025}.
In particular, it has been proposed that membraneless organelles, i.e.,
liquid-like intracellular condensates, may form by a mechanism involving fluids
with short-range attraction and long-range repulsion (SALR) rather than by
conventional liquid--liquid phase separation~\cite{Sweatman2019}. Monte Carlo
simulations of a SALR model showed that giant SALR clusters can form at a model
membrane surface even when they are unstable in the bulk, mimicking the puncta
formed at the cell membrane by the actin-regulatory protein WASP
(Wiskott--Aldrich syndrome protein) together with the adaptor protein Nck~\cite{sweatman2019assembly}.
Such multicomponent biological systems motivate the extension of SALR models
from one-component fluids to mixtures. At the single-component level, a SALR potential provides a coarse-grained,
effective description of the interactions between complex mesoscopic particles,
which are often mediated by a solvent. It is by now well established that such
interactions drive the spontaneous formation of periodically modulated phases,
i.e., microphase separation~\cite{ArcherWilding2007}. 
The resulting cluster, cylindrical, lamellar and gyroid-like morphologies, and the
thermodynamic conditions under which they arise, have been studied in
considerable detail by theory and simulations ~\cite{Ciach2008,Ciach2010,Ciach2013,zhuang:16:0,Edelmann2016,pini:17:0,ZhuangCharbonneau2016,Bretonnet2019,costa:23:0}.
The subject remains very active, with recent extensions to, for instance, self-propelled (active) SALR
suspensions~\cite{Serna2026}. Introducing a second, solvent-like species enlarges
this picture considerably and can give access to a broader family of self-assembled
morphologies depending on model parameters and thermodynamic conditions. 
For example, binary mixtures of oppositely charged colloids
self-assemble into ionic colloidal crystals~\cite{Leunissen2005}, while the
interplay of a short-range attraction and a long-range repulsion can arrest bulk
phase separation into equilibrium clusters and colloidal gels~\cite{Campbell2005}.
Understanding the
mechanisms of clustering in such mixtures is also important from an applied point
of view, as it may open additional possibilities for tailoring the properties of
novel nanomaterials. Moreover, cluster formation in mixtures of fluids with
competing interactions may take place in the region of the thermodynamic
parameters where it is absent for the one-component
fluid~\cite{Munao2022,PatsahanMeyraCiach2024}. The way in which the cross-interaction between the two
species controls clustering, secondary nucleation and microsegregation has been
addressed in several recent studies of binary SALR
mixtures~\cite{Tan2021,TanSweatman2024}, and a soft (penetrable)
cross-interaction has been argued to be, by itself, sufficient to drive microphase
separation in a binary colloidal system~\cite{Dhumal2026}. Investigating the
features brought about by a binary mixture in comparison with the one-component
case is thus of both fundamental and practical interest.

Given the large number of open questions concerning the properties of  mixtures with competing interactions, the influence of a spatial
confinement on such systems has been studied even
less.
At the same time, computer simulation data indicate the possibility of
non-trivial adsorption effects for such mixtures near a planar surface, for which
the density profile of the component with competing interactions differs
substantially from the one-component case~\cite{PatsahanMeyraCiach2022,LitniewskiCiach2021}. It has been shown that the adsorption
and the distribution of the particles near a wall depend essentially on the fluid
density and on the amplitude of the surface-particle interaction. Recent simulations further underline the richness of the structures formed by
confined systems with competing interactions. A mixture of nanoparticles with
competing interactions adsorbed onto a spherical colloid was found to
self-assemble into alternating layers, tori or spirals of the two
components~\cite{LitniewskiGozdzCiach2024}. Competing-interaction colloids
enclosed in spherical cavities form cluster, lamellar or perforated
morphologies depending on the curvature of the confining
wall~\cite{LiEtAl2024}, and confinement in ordered porous materials selects
different structures depending on the pore geometry~\cite{SernaNoyaGozdz2020}. 
In the related two-dimensional geometry, binary mixtures of particles with competing interactions
confined to a monolayer spontaneously organize into a variety of ordered
patterns~\cite{PatsahanMeyraCiach2024}.

In this work we take as our starting point the one-component soft-core
competing-interaction fluid at a hard wall, introduced
in~\cite{Kravtsiv2022}, and generalize it to a two-component mixture. Our main goal
is to understand how a solvent species modifies the structure of a
competing-interaction fluid near a hard surface. We pay particular
attention to the possibility of microsegregation of the two species and to the way in which
the concentrations and the temperature control it.

The theoretical description used here rests on a classical field-theoretical
formalism for inhomogeneous fluids. In this formalism the
partition function is represented as a functional integral over the density field,
whose saddle point yields the mean-field density profiles and, if required, may be
supplemented by Gaussian fluctuations. The scheme was originally devised for ionic
systems and electric double layers by Di~Caprio, Stafiej and
Badiali~\cite{DiCaprio1998,DiCaprio2001,DiCaprio2003,DiCaprioBadiali2008}. It was
subsequently transferred to fluids with Yukawa-type interactions confined by a
hard wall -- to a single-Yukawa fluid~\cite{HolovkoKravtsivSoviak2009,DiCaprio2011},
a two-Yukawa fluid~\cite{Kravtsiv2015} and an oscillating Yukawa
fluid~\cite{DiCaprio2016}, as well as to nematogenic fluids with an
orientation-dependent Maier-Saupe
interaction~\cite{HolovkoKravtsivDiCaprio2011,KravtsivHolovkoDiCaprio2013,HolovkoKravtsivDiCaprio2013,Holovko2016}
-- and, more recently, to the one-component fluid with a three-Yukawa (SALR)
competing interaction~\cite{Kravtsiv2022}, for which closed-form density profiles
were obtained. Here we extend this single-component treatment to a binary mixture
confined by a hard wall. In doing so we also
revisit the boundary conditions that fix the amplitudes of the linearized profiles:
in place of the contact theorem imposed on each Yukawa potential separately, we use
the exact conditions that follow from the continuity of the potentials at the
wall. In the mean-field
approximation this field-theoretical description is equivalent to the standard
mean-field density functional theory~\cite{Evans1979}. The mean-field predictions
are complemented by our own computer simulations. Related exact contact relations for
confining surfaces of non-planar geometry were recently established~\cite{Holovko2023}.

The remainder of the paper is structured as follows. In section~\ref{sec:model} we introduce the
model. In section~\ref{sec:field} we develop the field-theoretical mean-field
description of the mixture, derive the equations for the density profiles, and
formulate the contact theorem. Section~\ref{sec:numerical} presents the numerical
mean-field results for the bulk region and for a hard wall, together with a
comparison with molecular dynamics. In section~\ref{sec:analytical} we derive 
the analytical density profiles in the mesoscopically homogeneous region, 
bounded by the $\lambda$-surface, examine in detail the boundary conditions at the wall, and study the qualitative behaviour of the profiles. Conclusions are drawn in section~\ref{sec:concl}.

\section{Model}
\label{sec:model}

We consider a binary mixture of soft particles of two species. The interparticle
interaction of the main component (species~1) is modelled by a competing SALR
potential written as a sum of three Yukawa terms (3Y), while the interaction
between the solvent particles (species~2) and the cross interaction are described
by two Yukawa terms, one repulsive and one attractive (2Y):
\begin{align}
\label{pot11}
    \nu_{11}(r)&=\frac{A_{11}^{(1)}}{r}\,\exp\!\big(-\alpha_{11}^{(1)} r\big)
               +\frac{A_{11}^{(2)}}{r}\,\exp\!\big(-\alpha_{11}^{(2)} r\big)
               +\frac{A_{11}^{(3)}}{r}\,\exp\!\big(-\alpha_{11}^{(3)} r\big),\\
    \label{pot12}
    \nu_{12}(r)&=\frac{A_{12}^{(1)}}{r}\,\exp\!\big(-\alpha_{12}^{(1)} r\big)
               +\frac{A_{12}^{(2)}}{r}\,\exp\!\big(-\alpha_{12}^{(2)} r\big),\\
    \label{pot22}
    \nu_{22}(r)&=\frac{A_{22}^{(1)}}{r}\,\exp\!\big(-\alpha_{22}^{(1)} r\big)
               +\frac{A_{22}^{(2)}}{r}\,\exp\!\big(-\alpha_{22}^{(2)} r\big),
\end{align}
where $r$ is the distance between two particles, $A_{ij}^{(k)}$ are the amplitudes,
and $\alpha_{ij}^{(k)}$ the inverse ranges of the corresponding Yukawa terms;
throughout, the subscripts $i,j$ label the two species, while the superscript
$(k)$ numbers the individual Yukawa terms.
For the main component the amplitudes and ranges are tuned so that species~1
realizes the competing SALR form: $A_{11}^{(1)}>0$, $A_{11}^{(2)}<0$, $A_{11}^{(3)}>0$ with
$A_{11}^{(1)}>|A_{11}^{(2)}|>A_{11}^{(3)}$ and
$\alpha_{11}^{(1)}>\alpha_{11}^{(2)}>\alpha_{11}^{(3)}$, i.e., a short-range
attraction followed by a long-range repulsion. 
Since the amplitudes do not sum to zero, the potentials diverge at the origin
only as $\sum_k A_{ij}^{(k)}/r$. This weak, integrable Coulomb-like core is much
softer than the steep repulsion of hard-sphere or Lennard-Jones particles, so
that two particles can overlap partially. Such a soft core makes the model
suitable for coarse-grained descriptions of mesoscopic particles that can
interpenetrate, such as microgels, star and grafted polymers or globular
proteins, and also allows efficient molecular dynamics sampling.
The numerical values of the parameters used in this work
are collected in table~\ref{tab:param}.
The simplified model used in the analytical treatment
of section~\ref{sec:analytical} is defined by equation~(\ref{eq:simplified}).

\begin{table}[htb]
\caption{Parameters of the pair potentials~(\ref{pot11})--(\ref{pot22}).}
\label{tab:param}
\vspace{1ex}
\begin{center}
\begin{tabular}{lll}
\hline\noalign{\vskip 0.5ex}
	$A_{11}^{(1)}=150.6561472$  & $A_{11}^{(2)}=-122.6130119$  & $A_{11}^{(3)}=27.11810649$ \\
	$\alpha_{11}^{(1)}=1.92325375$  & $\alpha_{11}^{(2)}=1.261150$ & $\alpha_{11}^{(3)}=0.756690$ \\
	$A_{12}^{(1)}=413.5305236$    & $A_{12}^{(2)}=-275.6870157$  & \\
	$\alpha_{12}^{(1)}=3.405465108$    & $\alpha_{12}^{(2)}=3.0$ & \\
	$A_{22}^{(1)}=413.5305236$    & $A_{22}^{(2)}=-275.6870157$  & \\
	$\alpha_{22}^{(1)}=3.405465108$    & $\alpha_{22}^{(2)}=3.0$ & \\
\noalign{\vskip 0.5ex}\hline
\end{tabular}
\end{center}
\end{table}

Throughout this work, all energy quantities are in units of the depth of the attractive well of the main component, i.e., $|\nu_{11}(r_{\textrm{min}})|$, lengths are in
units of the distance $r_{0}$ at which $\nu_{11}(r_{0})=0$, densities are in units of
$r_{0}^{-3}$, and the reduced temperature is $k_{\textrm{B}}T/|\nu_{11}(r_{\textrm{min}})|$, where $k_B$ is the Boltzmann constant.
   
\begin{figure}[!htb]
\begin{center}
    \includegraphics[width=0.5\textwidth]{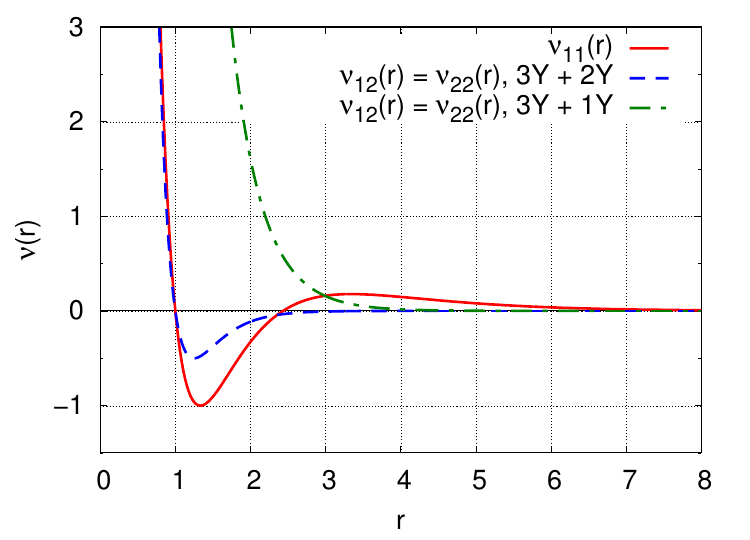}
    \caption{The interparticle pair potentials (\ref{pot11})--(\ref{pot22}) with the
    parameters of table~\ref{tab:param}: $\nu_{11}(r)$ (competing SALR, solid red
    line) and $\nu_{12}(r)=\nu_{22}(r)$ of the full $3\text{Y}{+}2\text{Y}$ model (dashed blue
    line). Both vanish at $r=1$, which serves as the soft-core particle diameter.
    The dash-dotted green line shows the purely repulsive
    $\nu_{12}(r)=\nu_{22}(r)$ of the simplified $3\text{Y}{+}1\text{Y}$
    model~(\ref{eq:simplified}) used in section~\ref{sec:analytical}.
    }
\label{fig:pot}
\end{center}
\end{figure}

To describe the presence of a surface, we introduce external potentials
$W_{1}(z)$ and $W_{2}(z)$ acting on the particles of species~1 and~2,
respectively, which describe a single hard wall or a slit-like pore bounded by
two hard walls,
\begin{align}
\label{wall}
    W_{i}(z)=\begin{cases}+\infty, & z<0,\\[0.3ex] 0, & z\geqslant0,\end{cases}
    \qquad
    \tilde{W}_{i}(z)=W_{i}(z)+W_{i}(L_{z}-z),\quad i=1,2,
\end{align}
where $z$ is the distance from the wall and $L_{z}$ is the width of the
slit-like pore (the second term describes the second wall; a single wall is
recovered as $L_{z}\to\infty$). The walls are thus purely hard: they exert no
force on either species and only exclude the particles from the region behind
the wall. The pair potentials~(\ref{pot11})--(\ref{pot22}) are shown in
figure~\ref{fig:pot}.

For the single-component version of this model, the extensive Monte Carlo and
molecular dynamics simulations reported in~\cite{Kravtsiv2022} show that the
competing potential~(\ref{pot11}) drives a rich sequence of self-assembled mesophases: 
the fluid is disordered at high temperature and, on cooling, organizes
into lamellar, gyroid, hexagonally packed cylindrical and cubic cluster (and void)
structures, the selected morphology depending on the density and the temperature.
These simulation snapshots are not reproduced here. The corresponding phase
diagram and illustrations of the mesophases can be found in~\cite{Kravtsiv2022}. 
Spatial confinement is known to reinforce this tendency to order,
the near-wall structuring surviving up to temperatures higher than those at which
the bulk fluid has already become homogeneous. In what follows we build on this
one-component picture and ask how it is modified once a solvent species is added.

\section{Field-theoretical mean-field description}
\label{sec:field}

\subsection{Hamiltonian and mean-field approximation}
\label{sec:hamiltonian}

In the field-theoretical formulation one starts from a Hamiltonian
$H[\rho(\mathbf{r})]$ that depends functionally on the density field
$\rho(\mathbf{r})$ and separates into an entropic (ideal-gas) part and an
interaction part. For a single-component fluid,
\begin{align}
\label{ham}
    \beta H[\rho(\mathbf{r}_{1})]&=
    \int \rho(\mathbf{r}_{1})\left(\ln\left[\rho(\mathbf{r}_{1})\Lambda^{3}\right]-1\right)\rd\mathbf{r}_{1}
    +\frac{\beta}{2}\int\nu(r_{12})\Big[\rho(\mathbf{r}_{1})\rho(\mathbf{r}_{2})
    -\rho(\mathbf{r}_{1})\delta(\mathbf{r}_{1}-\mathbf{r}_{2})\Big]\rd\mathbf{r}_{1}\rd\mathbf{r}_{2},
\end{align}
where $\beta=1/(k_{\textrm{B}}T)$ is the inverse temperature, $\rho(\mathbf{r})$ is
the number density and $\Lambda$ the de Broglie wavelength. Taking into account
the potentials~(\ref{pot11})--(\ref{pot22}), the Hamiltonian of the two-component
mixture reads
\begin{align}
\label{ham_w}
    \beta H[\rho_{1},\rho_{2}]&=
    \sum_{i=1}^{2}\int \rho_{i}(\mathbf{r}_{1})\left(\ln\left[\rho_{i}(\mathbf{r}_{1})\Lambda^{3}\right]-1\right)\rd\mathbf{r}_{1}\nonumber\\
    &+\frac{\beta}{2}\sum_{i,j=1}^{2}\int\nu_{ij}(r_{12})
    \Big[\rho_{i}(\mathbf{r}_{1})\rho_{j}(\mathbf{r}_{2})
    -\rho_{i}(\mathbf{r}_{1})\delta(\mathbf{r}_{1}-\mathbf{r}_{2})\delta_{ij}\Big]\rd\mathbf{r}_{1}\rd\mathbf{r}_{2},
\end{align}
where $\rho_{i}(\mathbf{r})$ denotes the density field of species~$i$. The partition
function and the Helmholtz free energy are given by
\begin{align}
\label{partition}
    Z\left[\rho\right]=\int\mathcal{D}\rho\,\exp\{-\beta H[\rho]\},\qquad
    \beta F=-\ln Z,
\end{align}
where the functional integration runs over all density distributions with a fixed
total number of particles of each species. In the mean-field approximation (MFA)
the free energy is obtained by minimizing the Hamiltonian with respect to the
density fields,
\begin{align}
\label{mfa}
    \frac{\delta\beta H}{\delta\rho_{1}}\bigg\vert_{\rho^{\textrm{MF}}}=\lambda_{1},\qquad
    \frac{\delta\beta H}{\delta\rho_{2}}\bigg\vert_{\rho^{\textrm{MF}}}=\lambda_{2},
\end{align}
$\lambda_{i}$ being Lagrange multipliers, so that $\beta F^{\textrm{MF}}=\beta
H^{\textrm{MF}}$.

\subsection{Bulk: partial pressures}
\label{sec:bulk}

In the spatially homogeneous (bulk) case the density fields are constant,
$\rho_{i}(\mathbf{r})=\rho_{ib}$, and the mean-field free energy becomes
\begin{align}
\label{Fbulk}
    \beta F^{\textrm{MF}}=\beta F_{\textrm{id}}
    +\frac{V}{2}\Big[\rho_{1b}^{2}\,\hat{\nu}_{11}
    +2\rho_{1b}\rho_{2b}\,\hat{\nu}_{12}
    +\rho_{2b}^{2}\,\hat{\nu}_{22}\Big],\qquad
    \hat{\nu}_{ij}=\beta\int\nu_{ij}(r)\,\rd\mathbf{r}
    =4\piup\beta\sum_{k}\frac{A_{ij}^{(k)}}{(\alpha_{ij}^{(k)})^{2}},
\end{align}
where $\beta F_{\textrm{id}}=\sum_{i}N_{i}(\ln[\rho_{ib}\Lambda^{3}]-1)$ is the
ideal-gas free energy, $N_{i}=\rho_{ib}V$, and $V$ is the volume. Introducing the
screening-like parameters
\begin{align}
\label{kappa}
    \varkappa_{ij}^{(k)2}=4\piup\beta\rho_{jb}A_{ij}^{(k)},
\end{align}
the bulk values of the mean-field potentials [defined in
equation~(\ref{Vdef}) below] are $\beta
V_{ijb}^{(k)}=\varkappa_{ij}^{(k)2}/(\alpha_{ij}^{(k)})^{2}$. Differentiation
of~(\ref{Fbulk}) yields the partial pressures,
\begin{align}
\label{P}
    \beta P_{1}^{\textrm{MF}}&=\rho_{1b}\left[1
    +\frac12\sum_{k=1}^{3}\frac{\varkappa_{11}^{(k)2}}{(\alpha_{11}^{(k)})^{2}}
    +\frac12\sum_{k=1}^{2}\frac{\varkappa_{12}^{(k)2}}{(\alpha_{12}^{(k)})^{2}}\right],\\
    \beta P_{2}^{\textrm{MF}}&=\rho_{2b}\left[1
    +\frac12\sum_{k=1}^{2}\frac{\varkappa_{22}^{(k)2}}{(\alpha_{22}^{(k)})^{2}}
    +\frac12\sum_{k=1}^{2}\frac{\varkappa_{21}^{(k)2}}{(\alpha_{21}^{(k)})^{2}}\right],
\end{align}
with the total pressure $\beta P^{\textrm{MF}}=\beta P_{1}^{\textrm{MF}}+\beta
P_{2}^{\textrm{MF}}$. These bulk relations are used below in the contact theorem
and in the comparison with the near-surface profiles.

\subsection{Density profiles near a surface and the contact theorem}
\label{sec:profiles}

In the presence of a surface, the minimization conditions~(\ref{mfa}) lead to the
following equations for the density profiles. Because the system remains uniform in
the two directions parallel to the wall, every position-dependent quantity is a
function of the single coordinate $z$ measured perpendicular to the surface:
\begin{align}
\label{euler1}
    \rho_{1}(z)&=\rho_{1b}\exp\left[-\beta\sum_{k=1}^{3}\!\big(V_{11}^{(k)}(z)-V_{11b}^{(k)}\big)
    -\beta\sum_{k=1}^{2}\!\big(V_{12}^{(k)}(z)-V_{12b}^{(k)}\big)-\beta\tilde{W}_{1}(z)\right],\\
\label{euler2}
    \rho_{2}(z)&=\rho_{2b}\exp\left[-\beta\sum_{k=1}^{2}\!\big(V_{22}^{(k)}(z)-V_{22b}^{(k)}\big)
    -\beta\sum_{k=1}^{2}\!\big(V_{21}^{(k)}(z)-V_{21b}^{(k)}\big)-\beta\tilde{W}_{2}(z)\right],
    \end{align}
where the mean-field potential created by species~$j$ and felt by species~$i$
through the $k$-th Yukawa term, and the corresponding field, are
\begin{align}
\label{Vdef}
    V_{ij}^{(k)}(\mathbf{r}_{1})=\int\rho_{j}(\mathbf{r}_{2})\,\frac{A_{ij}^{(k)}}{r_{12}}
    \exp(-\alpha_{ij}^{(k)}r_{12})\,\rd\mathbf{r}_{2},\qquad
    E_{ij}^{(k)}(\mathbf{r}_{1})=-\nabla V_{ij}^{(k)}(\mathbf{r}_{1}).
\end{align}
Acting on~(\ref{Vdef}) with the Laplacian and using the property of the Yukawa
potential, $(\Delta-(\alpha_{ij}^{(k)})^{2})V_{ij}^{(k)}=-4\piup
A_{ij}^{(k)}\rho_{j}$, together with the gradient of the logarithm of
equations~(\ref{euler1})--(\ref{euler2}), one obtains two coupled systems of
first-order differential equations for the unknown functions $\rho_{i}(z)$,
$V_{ij}^{(k)}(z)$ and $E_{ij}^{(k)}(z)$:
\begin{align}
\label{system1}
    \rho_{1}'(z)&=\rho_{1}(z)\left[\beta\sum_{k=1}^{3}E_{11}^{(k)}(z)+\beta\sum_{k=1}^{2}E_{12}^{(k)}(z)-\beta\tilde{W}_{1}'(z)\right],\nonumber\\
    V_{ij}^{(k)\prime}(z)&=-E_{ij}^{(k)}(z),\qquad
    E_{ij}^{(k)\prime}(z)=-(\alpha_{ij}^{(k)})^{2}V_{ij}^{(k)}(z)+\frac{\varkappa_{ij}^{(k)2}}{\beta\rho_{jb}}\rho_{j}(z),
\end{align}
for the pairs $ij=\{11,12\}$, and an analogous conjugate system for $\rho_{2}(z)$
with $ij=\{22,21\}$. This set of equations is solved numerically in
section~\ref{sec:numerical} and analytically in section~\ref{sec:analytical}.

The system~(\ref{system1}) admits a first integral, which provides an exact sum rule. 
For hard walls one can verify that
\begin{align}
\label{invariant}
    \frac{\rd}{\rd z}\Bigg[\rho_{1}(z)+\rho_{2}(z)
    &+\sum_{i=1}^{2}\sum_{k}\frac{\rho_{ib}}{2\varkappa_{ii}^{(k)2}}
    \Big\{\big(\alpha_{ii}^{(k)}\big)^{2}\big(\beta V_{ii}^{(k)}\big)^{2}
    -\big(\beta E_{ii}^{(k)}\big)^{2}\Big\}\nonumber\\
    &+\sum_{k}\frac{1}{4\piup\beta A_{12}^{(k)}}
    \Big\{\big(\alpha_{12}^{(k)}\big)^{2}\beta^{2}V_{12}^{(k)}V_{21}^{(k)}
    -\beta^{2}E_{12}^{(k)}E_{21}^{(k)}\Big\}\Bigg]=0,
\end{align}
where $1/(4\piup\beta A_{12}^{(k)})=\rho_{1b}/\varkappa_{21}^{(k)2}
=\rho_{2b}/\varkappa_{12}^{(k)2}$.

Two features distinguish~(\ref{invariant}) from its one-component
counterpart~\cite{Kravtsiv2022}, and both are essential. The densities enter
unscaled, and the cross fields appear through the bilinear products
$V_{12}^{(k)}V_{21}^{(k)}$ and $E_{12}^{(k)}E_{21}^{(k)}$ rather than through
separate squares. The reason is that $V_{12}^{(k)}$ is created by the solvent and
$V_{21}^{(k)}$ by the main component, so that the two are different functions of
$z$, and their contributions cancel only in the combination above due to the
identity
\begin{align}
\label{crossid}
    \rho_{1}\beta E_{12}^{(k)}+\rho_{2}\beta E_{21}^{(k)}
    =\frac{1}{4\piup\beta A_{12}^{(k)}}\frac{\rd}{\rd z}
    \Big[\beta^{2}E_{12}^{(k)}E_{21}^{(k)}
    -\big(\alpha_{12}^{(k)}\big)^{2}\beta^{2}V_{12}^{(k)}V_{21}^{(k)}\Big],
\end{align}
which follows from~(\ref{system1}). Replacing the densities by
$\rho_{i}/\rho_{ib}$ and the bilinear by separate squares, as would be natural by
analogy with the one-component case, leaves a residual
$\big(\sum_{k}\beta E_{12}^{(k)}-\sum_{k}\beta E_{21}^{(k)}\big)
\big(\rho_{1}/\rho_{1b}-\rho_{2}/\rho_{2b}\big)$, which does not vanish across the
interface.

Evaluating the conserved quantity in the bulk, where
$\rho_{i}(z)\to\rho_{ib}$, $E_{ij}^{(k)}\to0$ and $V_{ij}^{(k)}\to V_{ijb}^{(k)}$,
and at contact ($z=0_{+}$), and using~(\ref{P}), we obtain the contact theorem for
the two-component system:
\begin{align}
\label{contact}
    \beta P_{1}^{\textrm{MF}}+\beta P_{2}^{\textrm{MF}}=\rho_{1}(0_{+})+\rho_{2}(0_{+}),
\end{align}
which generalizes the one-component relation $\rho(0_{+})=\beta
P^{\textrm{MF}}$~\cite{Kravtsiv2022}. At contact every field term
in~(\ref{invariant}) vanishes, since the absence of fluid behind the wall makes
$\beta E_{ij}^{(k)}(0)=-\alpha_{ij}^{(k)}\beta V_{ij}^{(k)}(0)$. The same
conditions are used in section~\ref{sec:bc} to fix the amplitudes of the
linearized profiles, so that the contact theorem and the boundary
conditions~(\ref{bcfield}) are one and the same statement. 
The sum rule~(\ref{contact}) relates the \emph{total} contact density 
to the total mean-field pressure and is satisfied
exactly by the numerical solution. The species-resolved relations
$\beta P_{i}^{\textrm{MF}}=\rho_{i}(0_{+})$ do not follow from it and are not
imposed in the analytical treatment of section~\ref{sec:analytical}: there the
amplitudes are fixed by the exact conditions~(\ref{bcfield}), and the contact
density of each species follows from the solution.

\begin{figure}[!htb]
	\begin{center}
		\includegraphics[width=0.48\textwidth]{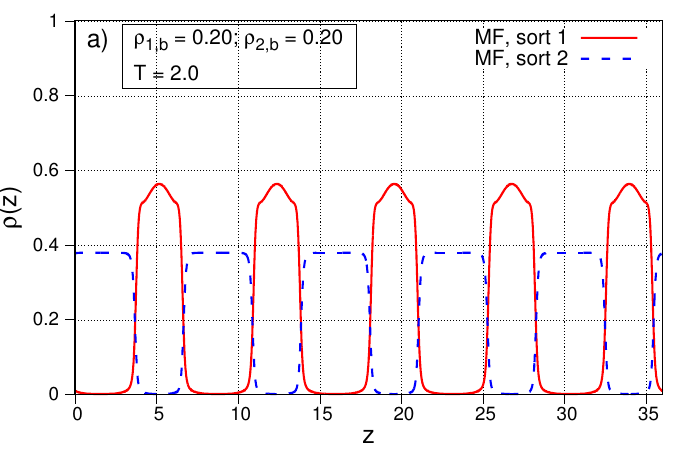}
		\includegraphics[width=0.48\textwidth]{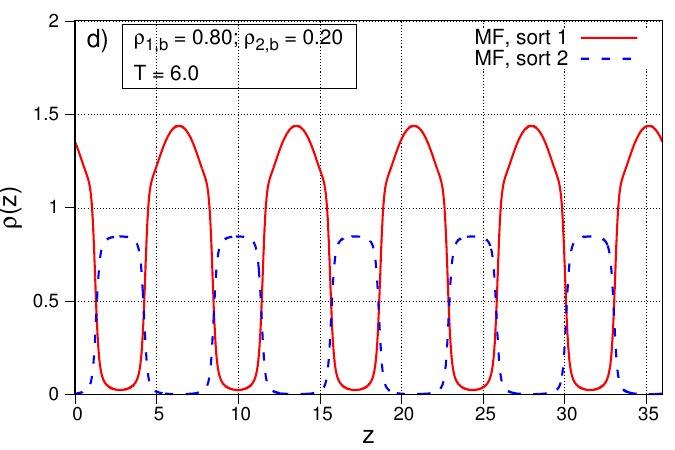} \\
		\includegraphics[width=0.48\textwidth]{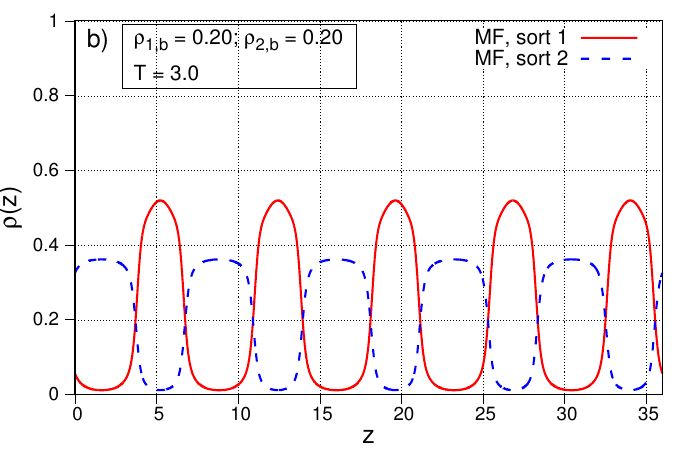}
		\includegraphics[width=0.48\textwidth]{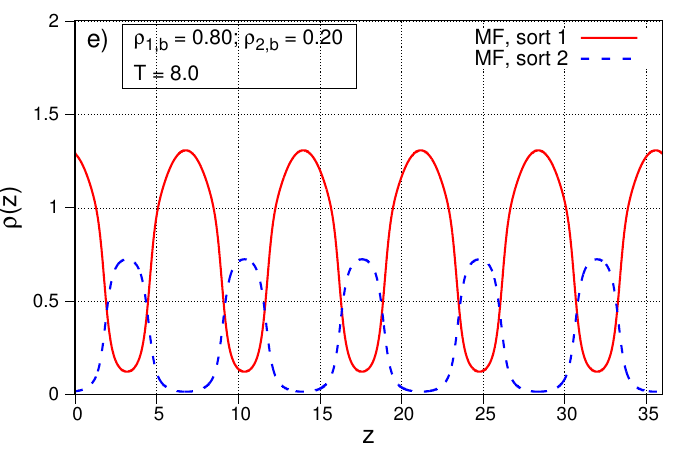} \\
		\includegraphics[width=0.48\textwidth]{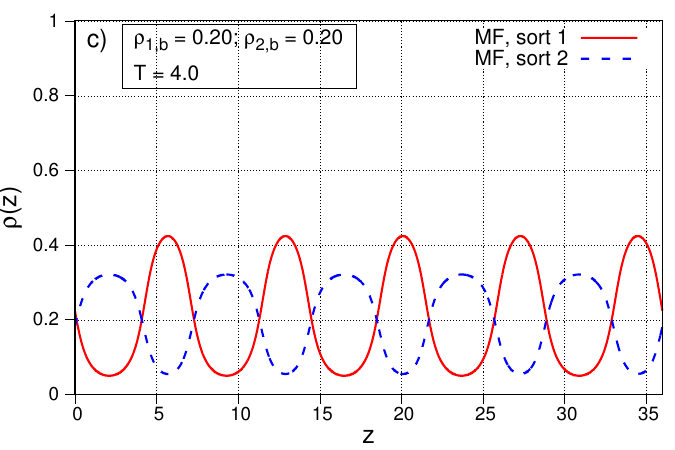}
		\includegraphics[width=0.48\textwidth]{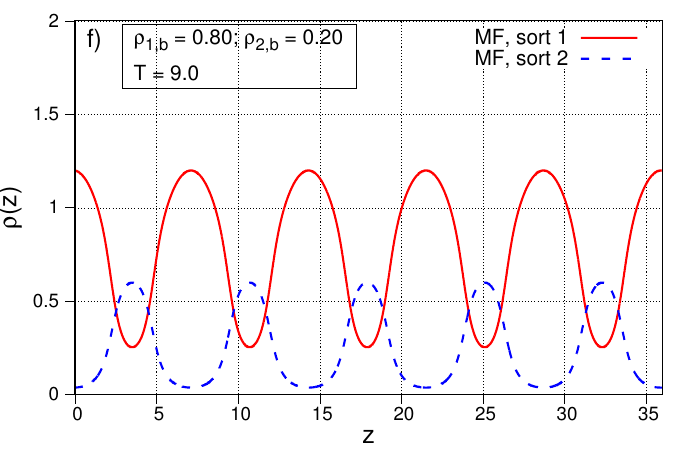}
		\caption{Bulk mean-field density profiles of the mixture at equal concentrations
			$\rho_{1b}=\rho_{2b}=0.20$ for (a)~$T=2$, (b)~$T=3$, (c)~$T=4$ 
            and at unequal concentrations $\rho_{1b}=0.80$, $\rho_{2b}=0.20$ for (d)~$T=6$, (e)~$T=8$, (f)~$T=9$.
            Solid red lines: species~1; dashed blue lines:
			species~2.}
		\label{fig:bulk}
	\end{center}
\end{figure}

\section{Numerical results in the mean-field approximation}
\label{sec:numerical}

\subsection{Bulk}
\label{sec:num_bulk}

\begin{figure}[!htb]
	\begin{center}
		\includegraphics[width=0.46\textwidth]{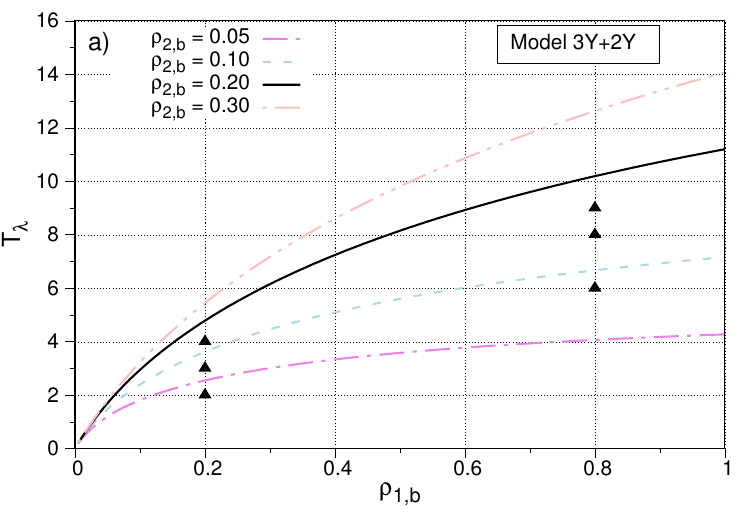}
		\includegraphics[width=0.48\textwidth]{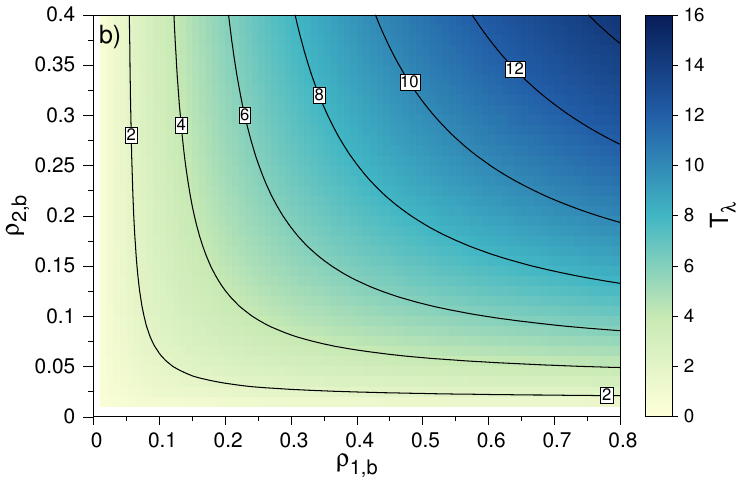}
		\caption{The $\lambda$-surface of the full $3\text{Y}{+}2\text{Y}$ model:
		the temperature $T_{\lambda}$ below which the bulk mixture ceases to be
		mesoscopically homogeneous and develops periodic density modulations.
		(a)~$T_{\lambda}$ as a function of the density of the main component $\rho_{1b}$
		at fixed solvent densities $\rho_{2b}=0.05$, $0.10$, $0.20$ and $0.30$;
		(b)~$T_{\lambda}$ over the $(\rho_{1b},\rho_{2b})$ plane (colour scale). The
		mixture is mesoscopically homogeneous above this surface and microsegregated
		below it.}
		\label{fig:lambda3Y2Y}
	\end{center}
\end{figure}

We first solve the mean-field equations for the mixture in the bulk, i.e., without walls, by imposing periodic boundary conditions on $\rho_1(z)$ and
$\rho_2(z)$ over a period $L_z$, and follow the evolution of the density profiles with temperature. At high temperatures both species are homogeneously
mixed. Upon cooling, the main component spontaneously develops periodic density modulations, and below a certain temperature the mixture separates into
alternating layers rich in species~1 and in species~2. Where the density of one species is close to its maximum, the density of the other is nearly zero. This
sequence is qualitatively similar to the one found in simulations of the one-component fluid~\cite{Kravtsiv2022} (see section~\ref{sec:model}).
The onset temperature of this microsegregation depends on the composition. 
For equal concentrations $\rho_{1b}=\rho_{2b}=0.20$ the modulations appear below $T\approx5$ (figure~\ref{fig:bulk}\,a--c), 
whereas for a larger fraction of the main component ($\rho_{1b}=0.80$, $\rho_{2b}=0.20$) they persist up to $T\approx10$ (figure~\ref{fig:bulk}\,d--f). 
In the mean-field approximation, the full extent of the modulated region is given 
by the $\lambda$-surface $T_{\lambda}(\rho_{1b},\rho_{2b})$ of the $3\text{Y}{+}2\text{Y}$ model shown in figure~\ref{fig:lambda3Y2Y}. 
Above this surface the homogeneous mixture is stable, while below it the mixture is microsegregated into alternating layers.

It is worth noting that, within the mean-field approximation, the one-component fluid with the potential (\ref{pot11}) shows no modulational instability at all, since $\hat\nu_{11}(q)>0$ for all $q$
(see section~\ref{sec:lin}). The microsegregation found here is therefore induced by the second component: the solvent shifts the onset of periodic ordering into a region where the pure competing interaction
fluid remains homogeneous at the mean-field level, whereas in simulations
the one-component fluid orders only at lower temperatures~\cite{Kravtsiv2022}.

\begin{figure}[!htb]
\begin{center}
{\color{black}
    \panel{0.30\textwidth}{0.5}{102}{a}{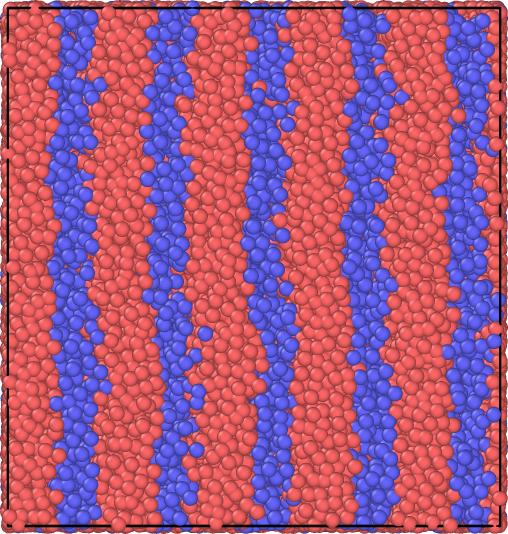}\hfill
    \panel{0.30\textwidth}{0.5}{102}{b}{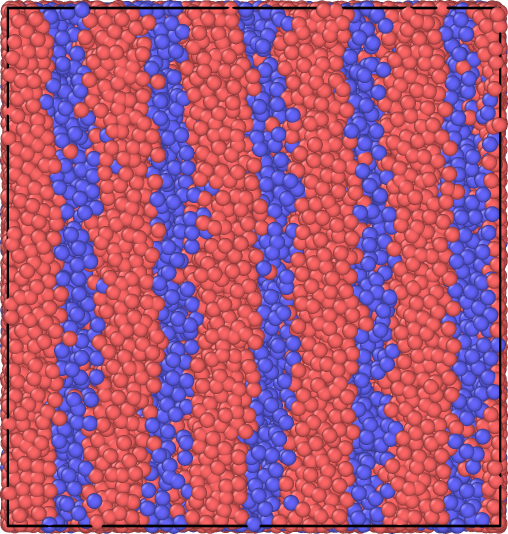}\hfill
    \panel{0.30\textwidth}{0.5}{102}{c}{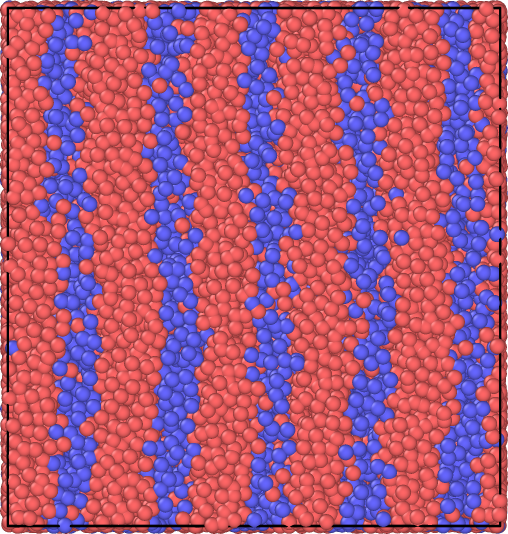}\\[2ex]
    \panel{0.30\textwidth}{0.5}{102}{d}{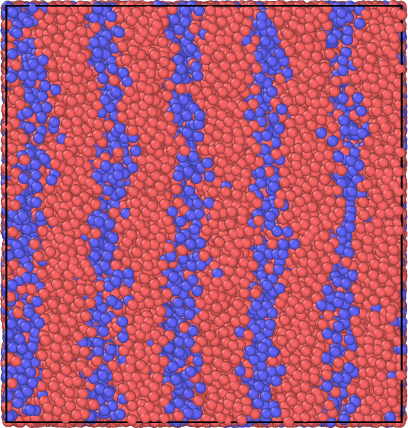}\hfill
    \panel{0.30\textwidth}{0.5}{102}{e}{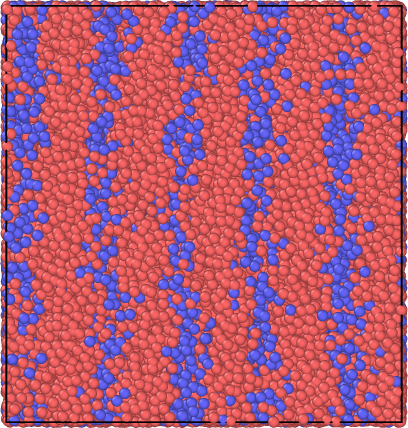}\hfill
    \panel{0.30\textwidth}{0.5}{102}{f}{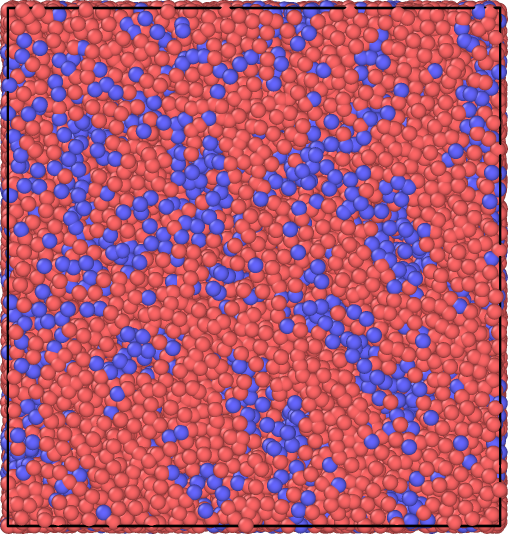}
    \caption{Equilibrated configurations from molecular dynamics for the bulk mixture
    at $\rho_{1b}=0.8$, $\rho_{2b}=0.2$: (a)~$T=2.0$, (b)~$T=2.5$, (c)~$T=3.0$,
    (d)~$T=3.5$, (e)~$T=4.0$ and (f)~$T=4.5$. The lamellar structure formed at low
    temperature becomes progressively more diffuse on heating and is destroyed between
    $T=4.0$ and $T=4.5$.}
\label{fig:snap}
}
\end{center}
\end{figure}

These mean-field results can be compared with direct simulations of the same model. 
Figure~\ref{fig:snap} shows equilibrated configurations from molecular
dynamics in the bulk at $\rho_{1b}=0.80$, $\rho_{2b}=0.20$, the composition of
figure~\ref{fig:bulk}\,d--f, over the range $T=2.0$--$4.5$. At $T=2.0$ the mixture has segregated 
into a clear lamellar structure of alternating slabs of species~1 and species~2. 
The same structure survives at $T=2.5$--$3.5$, with progressively more diffuse interfaces between the slabs. 
At $T=4.0$ the layers begin to lose their shape, and at $T=4.5$ they are destroyed altogether. 
The onset of microsegregation in the simulated system thus lies between $T=4.0$ and
$T=4.5$, well below the mean-field estimate $T\approx10$ (see section~\ref{sec:num_md}).

\subsection{Hard wall}
\label{sec:num_hard}
\begin{figure}[!htb]
\begin{center}
	\panel{0.46\textwidth}{0.5}{63}{a}{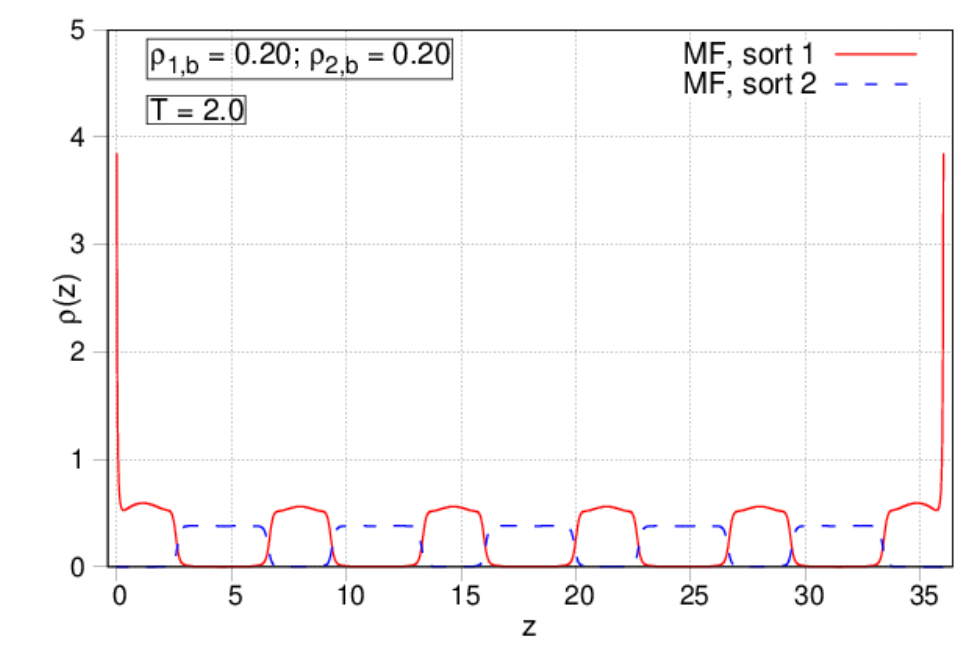}\hfill
	\panel{0.46\textwidth}{0.5}{63}{b}{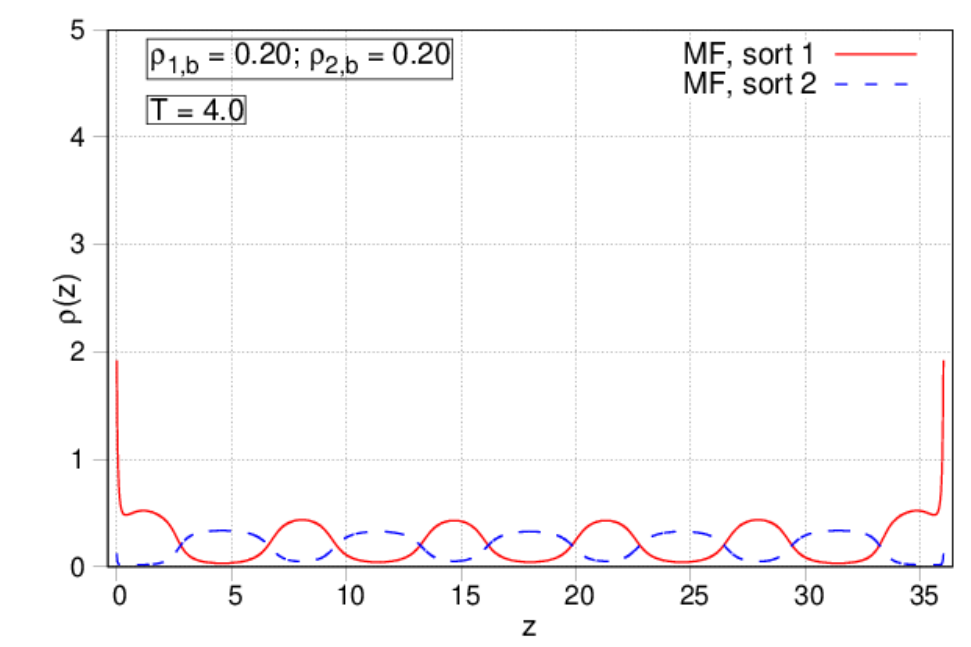}\\[3ex]
	\panel{0.46\textwidth}{0.5}{63}{c}{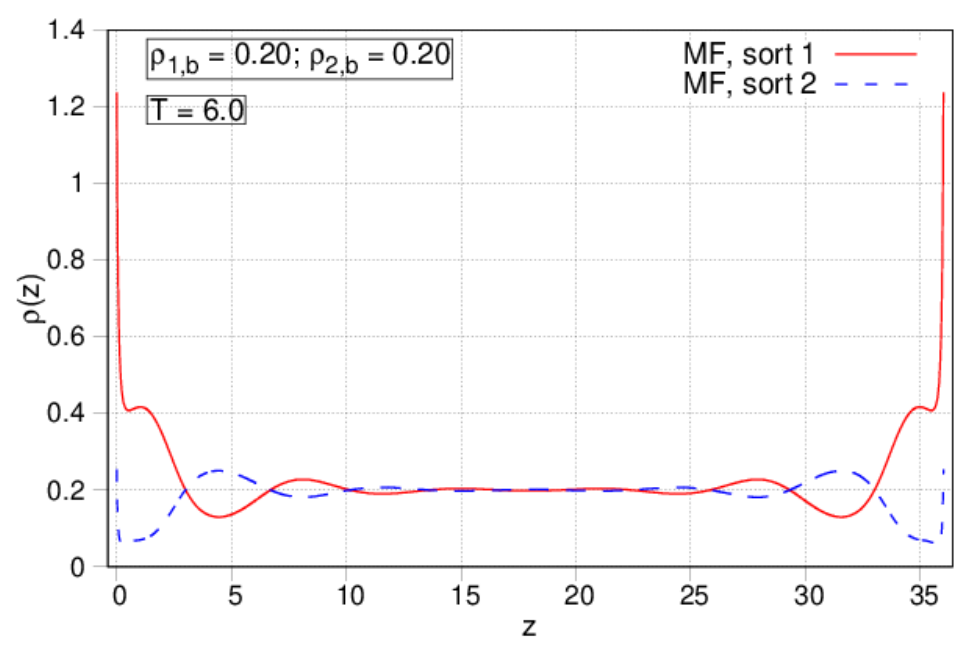}\hfill
	\panel{0.46\textwidth}{0.5}{63}{d}{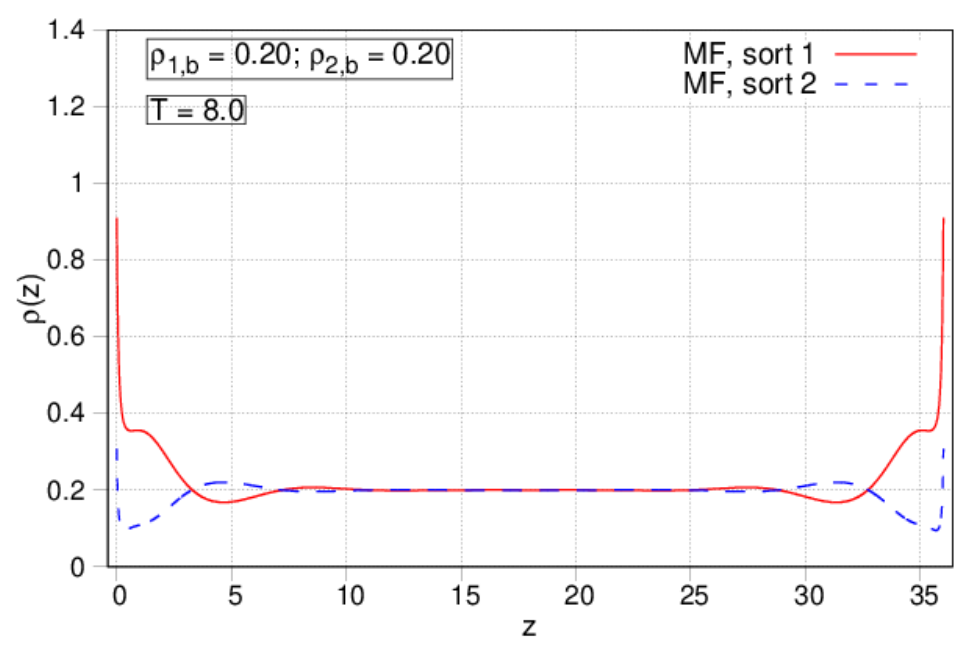}
	\caption{Density profiles of the binary mixture between two hard walls 
    for bulk densities $\rho_{1b}=\rho_{2b}=0.20$ and temperatures (a)~$T=2$, (b)~$T=4$, (c)~$T=6$ and (d)~$T=8$. Solid red lines: species~1; dashed blue lines: species~2.}
	\label{fig:attr1}
\end{center}
\end{figure}

\begin{figure}[!htb]
\begin{center}
    \panel{0.46\textwidth}{0.5}{63}{a}{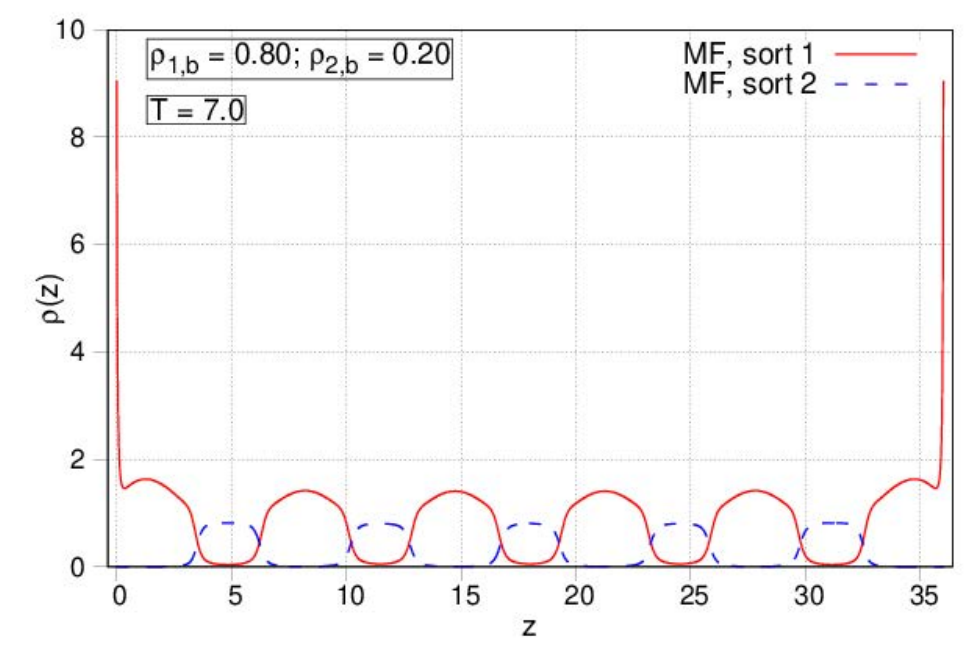}\hfill
    \panel{0.46\textwidth}{0.5}{63}{b}{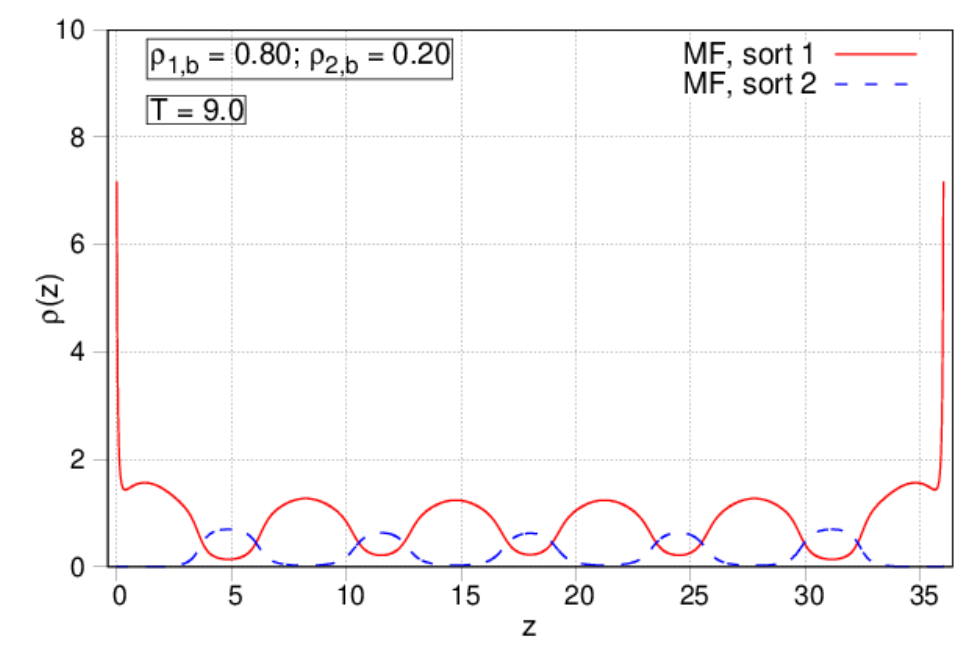}\\[3ex]
    \panel{0.46\textwidth}{0.5}{63}{c}{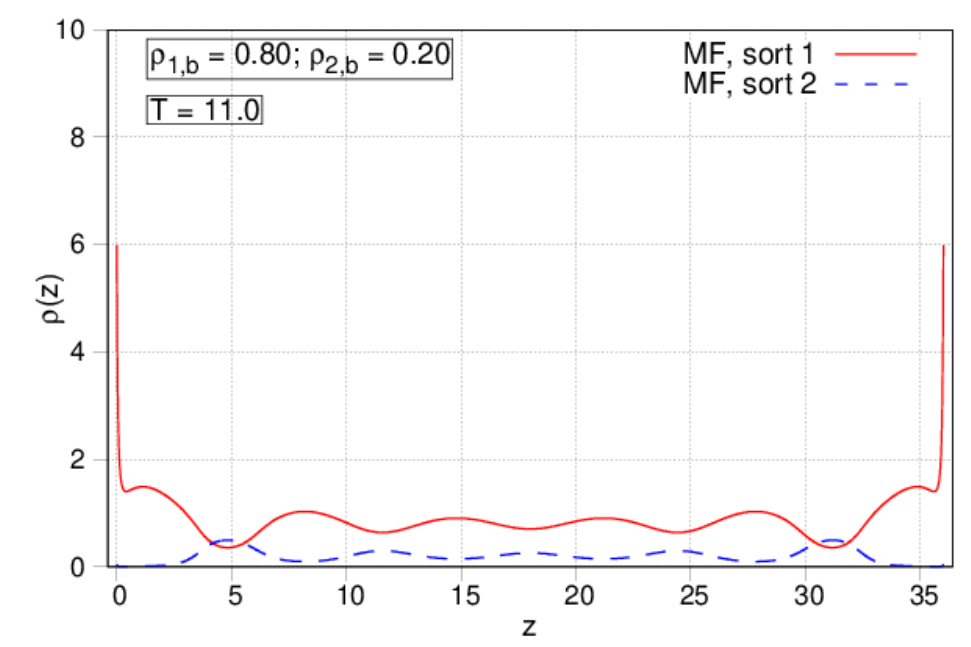}\hfill
    \panel{0.46\textwidth}{0.5}{63}{d}{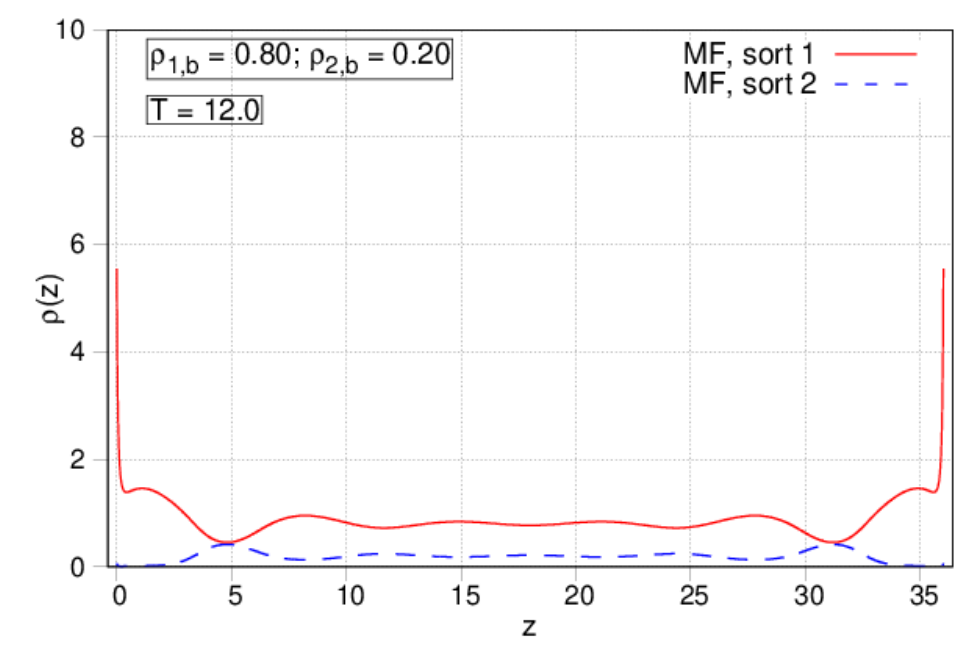}
    \caption{Density profiles of the binary mixture between two hard walls for
    bulk densities $\rho_{1b}=0.80$, $\rho_{2b}=0.20$ and temperatures
    (a)~$T=7$, (b)~$T=9$, (c)~$T=11$ and (d)~$T=12$. Solid red lines: species~1;
    dashed blue lines: species~2.}
\label{fig:hardwalls}
\end{center}
\end{figure}

Figures~\ref{fig:attr1} and~\ref{fig:hardwalls} show the
density profiles of the mixture confined between two hard walls at several
temperatures: figure~\ref{fig:attr1} for equal bulk concentrations
$\rho_{1b}=\rho_{2b}=0.20$ ($T=2$--$8$), and figure~\ref{fig:hardwalls} for
$\rho_{1b}=0.80$, $\rho_{2b}=0.20$. For the latter composition, at the lowest temperatures ($T\lesssim9$) 
the two components form alternating layers in which the concentration of one species is close to zero
while that of the other reaches its maximal value. These oscillations do not
decay towards the middle of the slit. At higher temperatures, the mixture 
becomes homogeneous far from the wall where the densities of the species reach their bulk values.
The ordering is destroyed across the slit only at $T\approx12$, i.e., at a higher temperature
than in the corresponding bulk system ($T\approx10$). The confining wall thus
promotes the ordering -- it acts as a nucleation site for the near-surface
structure, which persists at all temperatures considered. Near the surface the
structure of the mixture differs from the bulk in all cases: the main component is
adsorbed, its contact value substantially exceeding the bulk one, and a layer a
few effective diameters wide is formed in which the solvent is almost absent. The
behaviour of the main component near the wall is reminiscent of the one-component
case studied in~\cite{Kravtsiv2015,Kravtsiv2022}, in particular the two-peak
structure that develops immediately in front of the wall, whereas the solvent is
expelled from the near-surface region by the main component at all temperatures
considered. A qualitatively similar near-surface structure, consisting of separate,
partially mixed and fully mixed layers with a contact film of the main
component, was observed in molecular simulations of a related self-assembling
binary mixture near a wall~\cite{PatsahanMeyraCiach2022}.

\subsection{Comparison with computer simulations}
\label{sec:num_md}

The theoretical predictions have been verified and complemented by molecular
dynamics simulations of the mixture of species~1 and~2 in the canonical
ensemble~\cite{allen2017computer}, carried out with the LAMMPS
package~\cite{thompson2022lammps}, in which the pair
interactions~(\ref{pot11})--(\ref{pot22}) were implemented specifically for this
purpose. The bulk runs used a cubic cell of side $L=36$ with periodic boundary
conditions in all three directions, containing $N_1=\rho_{1b}L^3$ and
$N_2=\rho_{2b}L^3$ particles of the two species. In the slit geometry a cell of
the same size was used, with the walls placed at $z=0$ and $z=L_z=L$ and
implemented as reflecting hard walls, and periodic boundary conditions were
applied only in the $x$ and $y$ directions. All particles have the same mass
$m=1$, and time is measured in units of $\tau=r_0\sqrt{m/|\nu_{11}(r_{\min})|}$.
The equations of motion were integrated with the velocity Verlet algorithm at a
time step $\delta t=0.01\tau$. The temperature was controlled by a
Nos\'e--Hoover thermostat with a relaxation time of $1.0\tau$, and the pair
potentials were truncated at $r_{c}=12.0$. Starting from a random placement of
particles of both species, each system was equilibrated over $10^5$ steps, after
which the density profiles of both species were accumulated over a further
$10^5$ steps, with configurations sampled every $100$ steps.

For the bulk mixture, the simulation results are presented only as the
representative snapshots of figure~\ref{fig:snap}. A quantitative comparison is
made for the $3\text{Y}{+}2\text{Y}$ mixture in the slit geometry. As noted in
section~\ref{sec:num_bulk}, the simulations place the onset of microsegregation
in the bulk mixture with $\rho_{1b}=0.80$, $\rho_{2b}=0.20$ between $T=4.0$ and
$T=4.5$, whereas the mean-field theory gives $T\approx10$. This reflects a
general feature of the mean-field approximation, which systematically
overestimates the temperature at which microsegregation sets in, so that the
$\lambda$-surface obtained from the theory lies at higher temperatures than the
boundary of the modulated region observed in simulation. A comparison of the
two profiles at one and the same temperature could therefore place the system
on opposite sides of this boundary and would not be meaningful. Instead, the
mean-field and simulation profiles are compared at shifted temperatures, the
simulation temperature being taken lower than the mean-field one. The shift is
larger for the denser mixture, in line with the larger mean-field overestimate
found there.

\begin{figure}[!htb]
	\begin{center}
		\includegraphics[width=0.48\textwidth]{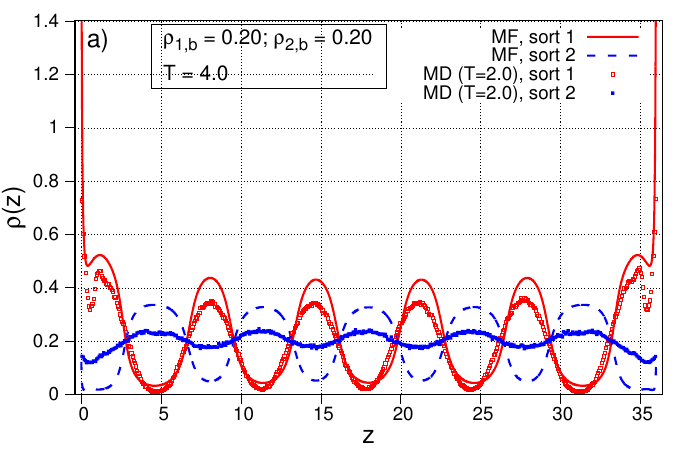}
		\includegraphics[width=0.48\textwidth]{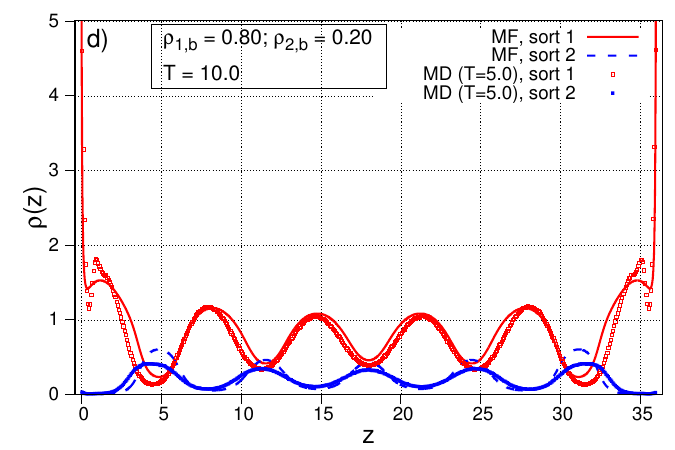} \\
		\includegraphics[width=0.48\textwidth]{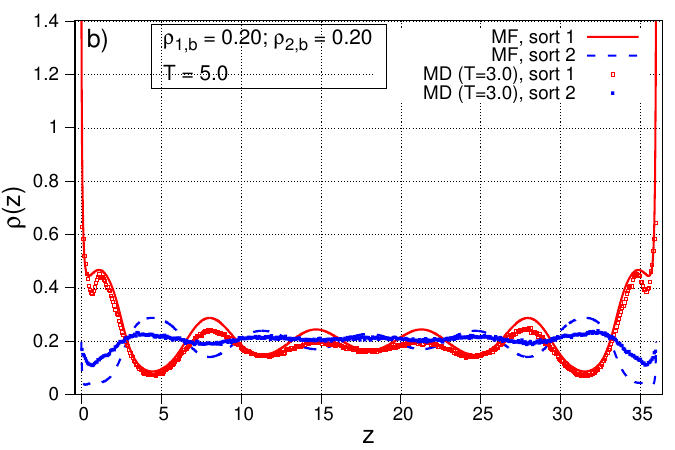}
		\includegraphics[width=0.48\textwidth]{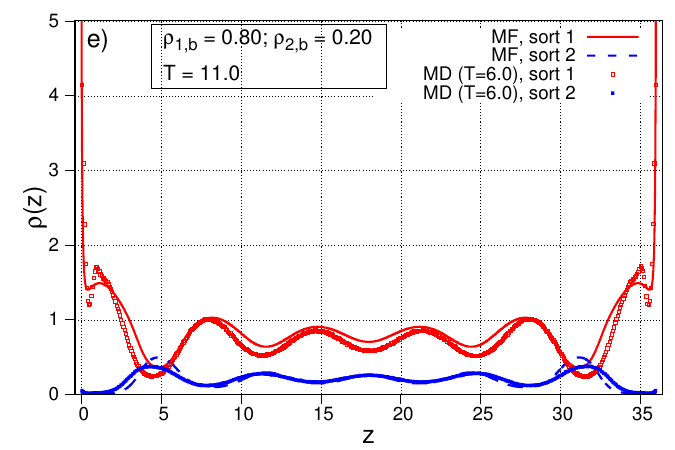} \\
		\includegraphics[width=0.48\textwidth]{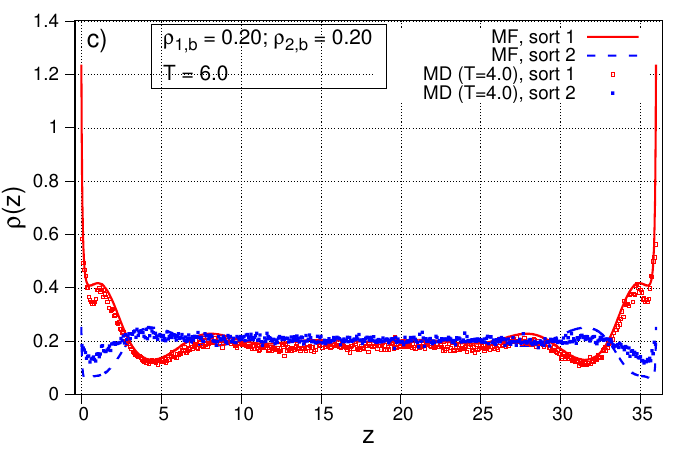}
		\includegraphics[width=0.48\textwidth]{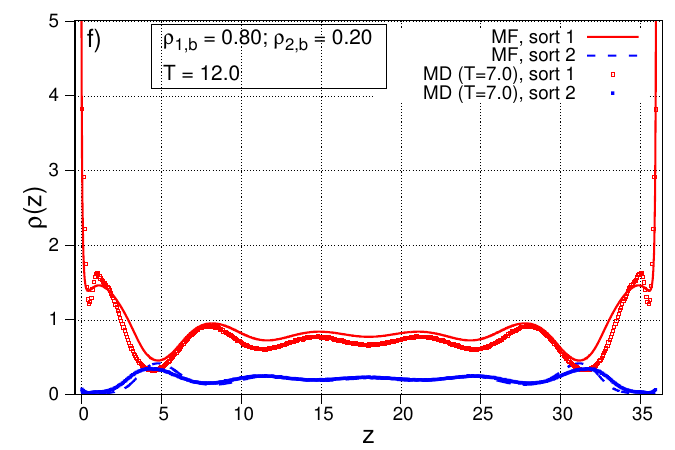}
		\caption{Density profiles from the mean-field theory (MF, lines) and from molecular
		dynamics simulations (MD, symbols) for $\rho_{1b}=\rho_{2b}=0.20$: (a)~MF at $T=4$
		versus MD at $T=2$, (b)~MF at $T=5$ versus MD at $T=3$, and (c)~MF at
		$T=6$ versus MD at $T=4$ and for $\rho_{1b}=0.80$, $\rho_{2b}=0.20$: (d)~MF at $T=10$
		versus MD at $T=5$, (e)~MF at $T=11$ versus MD at $T=6$, and (f)~MF at
		$T=12$ versus MD at $T=7$}. The simulation temperature is taken lower than
		the mean-field one to compensate for the mean-field overestimation of the onset temperature of microsegregation (see text). Red: species~1; blue: species~2.
		\label{fig:md}
	\end{center}
\end{figure}

Figure~\ref{fig:md} shows three such pairs for each of the two mixtures
confined by hard walls, with equal densities $\rho_{1b}=\rho_{2b}=0.20$
(figure~\ref{fig:md}\,a--c) and with $\rho_{1b}=0.80$, $\rho_{2b}=0.20$
(figure~\ref{fig:md}\,d--f). For the first mixture, the mean-field profiles at
$T=4$ are compared with the simulation data at $T=2$ in
figure~\ref{fig:md}\,a. Both species show undamped oscillations across the
whole slit, and the theory reproduces their wavelength and the positions of the
maxima and minima. The agreement is close for the competing-interaction
component, whereas for the solvent the theory considerably overestimates the
amplitude of the oscillations, which are only weak in simulation. As the
temperature is raised (mean field at $T=5$ versus simulation at $T=3$, and mean
field at $T=6$ versus simulation at $T=4$), the profiles become only weakly
structured except near the walls, where theory and simulation agree well for
both components (figure~\ref{fig:md}\,b,c).

The agreement is even better for the denser mixture,
$\rho_{1b}=0.80$, $\rho_{2b}=0.20$ (figure~\ref{fig:md}\,d--f). At the lowest
pair of temperatures (mean field at $T=10$ versus simulation at $T=5$) both
descriptions show alternating layers of the two species that do not decay
towards the middle of the slit. The positions and heights of the maxima of the
main component practically coincide, and the solvent is concentrated in the
minima of the main-component profile in both cases. Near each wall, theory and
simulation consistently show a strong adsorption of the main component with a
large contact value, the two-peak structure in front of the wall, and a layer
a few diameters wide from which the solvent is expelled. The mean-field theory
only slightly overestimates the height of the first solvent maximum. On
heating (mean field at $T=11$ and $12$ versus simulation at $T=6$ and $7$), the
oscillations in the middle of the slit are progressively damped in both
descriptions at a similar rate, while the near-wall structure remains almost
unchanged. Thus, once the shift of the onset temperature is taken into account,
the mean-field theory provides a quantitatively reasonable description of the
simulated profiles, particularly for the main component and for the denser
mixture.

\section{Analytical density profiles in the mesoscopically homogeneous region}
\label{sec:analytical}

Two restrictions underlie the closed-form results of this section, and we state
them explicitly at the outset. 
First, the analytical treatment is confined to the
\emph{mesoscopically homogeneous} part of the phase diagram, i.e., to
temperatures above the $\lambda$-surface, $T>T_\lambda$, where the near--wall
modulations decay into the bulk. There, sufficiently far from $T_\lambda$, the
deviations of the profiles from their bulk values are small and the linearization
of equations~(\ref{euler1})--(\ref{euler2}) is justified. 
The microsegregated regime below $T_\lambda$ is accessible only through the numerical solution of
section~\ref{sec:numerical}.
Second, to obtain the profiles in radicals we adopt a \emph{simplified} version
of the model in which the solvent--solvent and cross interactions are reduced to
a single repulsive Yukawa term identical to the leading term of $\nu_{11}$,
\begin{equation}
    A_{12}^{(1)}=A_{22}^{(1)}=A_{11}^{(1)},\qquad
    \alpha_{12}^{(1)}=\alpha_{22}^{(1)}=\alpha_{11}^{(1)},\qquad
    A_{12}^{(2)}=A_{22}^{(2)}=0,
    \label{eq:simplified}
\end{equation}
cf.\ equations~(\ref{pot12})--(\ref{pot22}). For the full $3\text{Y}{+}2\text{Y}$ model the
characteristic equation is of seventh degree, cannot in general be solved in
radicals, and must be treated numerically.
Following the shorthand of
section~\ref{sec:model}, we refer to the full model as the $3\text{Y}{+}2\text{Y}$ model and
to this reduced version as the $3\text{Y}{+}1\text{Y}$ model, after the number of Yukawa terms
in $\nu_{11}$ and in $\nu_{12}=\nu_{22}$ respectively. Under these two restrictions the problem becomes
analytically tractable. The scheme presented below extends to the binary mixture
the one-component derivation of~\cite{Kravtsiv2022}. Here we outline the main steps
and quote the final results, while the details of the calculation are given in
appendix~\ref{app:derivation}.

\subsection{Linearization and the eigenvalue problem}
\label{sec:lin}

In the mesoscopically homogeneous region the density profiles can be obtained in
closed form. The route to this is to linearize the
exponentials in equations~(\ref{euler1})-(\ref{euler2}),
\begin{align}
\label{lin}
    \rho_{1}(z)&=\rho_{1b}\Big[1-\beta\sum_{k=1}^{3}\!\big(V_{11}^{(k)}(z)-V_{11b}^{(k)}\big)
    -\beta\sum_{k=1}^{2}\!\big(V_{12}^{(k)}(z)-V_{12b}^{(k)}\big)\Big],\nonumber\\
    \rho_{2}(z)&=\rho_{2b}\Big[1-\beta\sum_{k=1}^{2}\!\big(V_{22}^{(k)}(z)-V_{22b}^{(k)}\big)
    -\beta\sum_{k=1}^{2}\!\big(V_{21}^{(k)}(z)-V_{21b}^{(k)}\big)\Big],
\end{align}
which turns the system~(\ref{system1}) into a set of linear equations for the
fields $E_{ij}^{(k)}(z)$. Eliminating the densities and potentials, one arrives at
a system of second-order differential  equations that can be written in matrix form,
\begin{align}
\label{matrix}
    \mathbf{E}''=\mathbf{A}\,\mathbf{E},
\end{align}
where $\mathbf{E}$ collects the seven independent fields $E_{11}^{(1,2,3)}$,
$E_{12}^{(1,2)}$, $E_{21}^{(1,2)}$, and $\mathbf{A}$ is a $7\times7$ matrix built from the parameters
$\alpha_{ij}^{(k)}$ and $\varkappa_{ij}^{(k)}$. The decay coefficients  of the profiles
are the square roots of the eigenvalues $\Lambda_{n}$ of $\mathbf{A}$, i.e., the
roots of the characteristic septic polynomial
$\det(\mathbf{A}-\Lambda\mathbf{I})=0$,
\begin{align}
    \label{quintic}
    \Lambda^{7}+a_{6}\Lambda^{6}+a_{5}\Lambda^{5}+a_{4}\Lambda^{4}
    +a_{3}\Lambda^{3}+a_{2}\Lambda^{2}+a_{1}\Lambda+a_{0}=0.
\end{align}

The septic equation~(\ref{quintic}) cannot be solved in radicals in the general
case. However, it reduces to a cubic equation in the eigenvalue $\Lambda$
(equivalently, a bicubic equation in the decay coefficient
$\lambda=\sqrt{\Lambda}$) if three conditions are met simultaneously:
(i)~the solvent--solvent and cross interactions are purely repulsive,
$A_{12}^{(2)}=A_{22}^{(2)}=0$; (ii)~the range of the remaining Yukawa term
coincides with that of the leading term of the main component,
$\alpha_{1}\equiv\alpha_{11}^{(1)}=\alpha_{12}^{(1)}=\alpha_{22}^{(1)}$; and
(iii)~so does its amplitude, $A_{12}^{(1)}=A_{22}^{(1)}=A_{11}^{(1)}$. Together
they define the simplified model~(\ref{eq:simplified}) treated in what follows;
without condition~(iii) the equation would be quartic. Under these conditions
$V_{21}^{(1)}=V_{11}^{(1)}$ and $V_{22}^{(1)}=V_{12}^{(1)}$, i.e., both species
feel the leading Yukawa term through the total density $\rho_1+\rho_2$.
Introducing the superposition of the two fields with the common range,
$\tilde{E}(z)=E_{11}^{(1)}(z)+E_{12}^{(1)}(z)$, that is the sum of the two
fields of range $\alpha_{1}$ felt by the main component, the problem reduces to
a closed system of three second-order equations for
$\mathbf{E}=(\tilde{E},E_{12},E_{13})^{\textrm{T}}$, with
\begin{align}
\label{matrix3}
    \mathbf{A}=
    \begin{pmatrix}
    \alpha_{1}^{2}+\varkappa_{11}^{2}+\varkappa_{\mathrm s}^{2}& \varkappa_{11}^{2} & \varkappa_{11}^{2}\\[0.6ex]
    \varkappa_{12}^{2} & \alpha_{2}^{2}+\varkappa_{12}^{2} & \varkappa_{12}^{2}\\[0.6ex]
    \varkappa_{13}^{2} & \varkappa_{13}^{2} & \alpha_{3}^{2}+\varkappa_{13}^{2}
    \end{pmatrix},
\end{align}
where we abbreviated $\varkappa_{1k}^2\equiv\varkappa_{11}^{(k)2}$ and
$\alpha_k\equiv\alpha_{11}^{(k)}$ ($k=1,2,3$), and
$\varkappa_{\mathrm s}^2\equiv4\pi\beta\rho_{2b}A_{11}^{(1)}
=(\rho_{2b}/\rho_{1b})\varkappa_{11}^{(1)2}$ is the coupling through which the
solvent enters the $(1,1)$ entry of $\mathbf A$.
The eigenvalues now solve the cubic equation
\begin{align}
\label{cubic}
    \Lambda^{3}-b\Lambda^{2}+c\Lambda-d=0,
\end{align}
with
\begin{align}
\label{coefs}
    b&=\tilde{\varkappa}_{11}^{2}+\tilde{\varkappa}_{12}^{2}+\tilde{\varkappa}_{13}^{2},\qquad
    c=\tilde{\varkappa}_{11}^{2}\tilde{\varkappa}_{12}^{2}+\tilde{\varkappa}_{11}^{2}\tilde{\varkappa}_{13}^{2}
    +\tilde{\varkappa}_{12}^{2}\tilde{\varkappa}_{13}^{2}
    -\varkappa_{11}^{2}\varkappa_{12}^{2}-\varkappa_{11}^{2}\varkappa_{13}^{2}-\varkappa_{12}^{2}\varkappa_{13}^{2},\nonumber\\
    d&=\tilde{\varkappa}_{11}^{2}\big(\tilde{\varkappa}_{12}^{2}\tilde{\varkappa}_{13}^{2}-\varkappa_{12}^{2}\varkappa_{13}^{2}\big)
    -\varkappa_{11}^{2}\varkappa_{12}^{2}\tilde{\varkappa}_{13}^{2}
    -\varkappa_{11}^{2}\varkappa_{13}^{2}\tilde{\varkappa}_{12}^{2}
    +2\varkappa_{11}^{2}\varkappa_{12}^{2}\varkappa_{13}^{2},
\end{align}
where $\tilde{\varkappa}_{11}^{2}=\varkappa_{11}^{2}+\varkappa_{\mathrm s}^{2}+\alpha_{1}^{2}$,
$\tilde{\varkappa}_{12}^{2}=\varkappa_{12}^{2}+\alpha_{2}^{2}$,
$\tilde{\varkappa}_{13}^{2}=\varkappa_{13}^{2}+\alpha_{3}^{2}$. For the range of
parameters considered here, equation~(\ref{cubic}) yields a single real solution
$\Lambda_{0}=\lambda_{0}^{2}$ together with a complex-conjugate pair
$\Lambda_{1,2}=\lambda_{1,2}^{2}$, $\lambda_{1,2}=\lambda\pm\ri\mu$, all following
in closed form from the standard formulae for the roots of a cubic equation. It is this
reduction that makes the linearized theory fully analytic, and it is the reason we
use the $3\text{Y}{+}1\text{Y}$ model to test the analytical description against the numerical
mean-field solution. For the full $3\text{Y}{+}2\text{Y}$ model the septic equation~(\ref{quintic})
has to be solved numerically. Its linearized profiles at hard and attractive walls
will be considered elsewhere. The
general solution for each field is a superposition of a monotonically decaying
term and an exponentially damped oscillation,
\begin{align}
\label{Esol}
    \tilde{E}(z)=c_{10}\re^{-\lambda_{0}z}
    +C_{1}\re^{(-\lambda+\ri\mu)z}+C_{2}\re^{(-\lambda-\ri\mu)z},
\end{align}
where only the terms decaying at $z\to\infty$ are retained, and $C_{1}$, $C_{2}$
are complex conjugate. Analogous expressions hold for $E_{12}(z)$ and $E_{13}(z)$. 
Since the system is linear and homogeneous, all coefficients can be expressed through a single set
of unknowns using the eigenvector relations.

\begin{figure}[htb]
\begin{center}
{
    \includegraphics[width=0.46\textwidth]{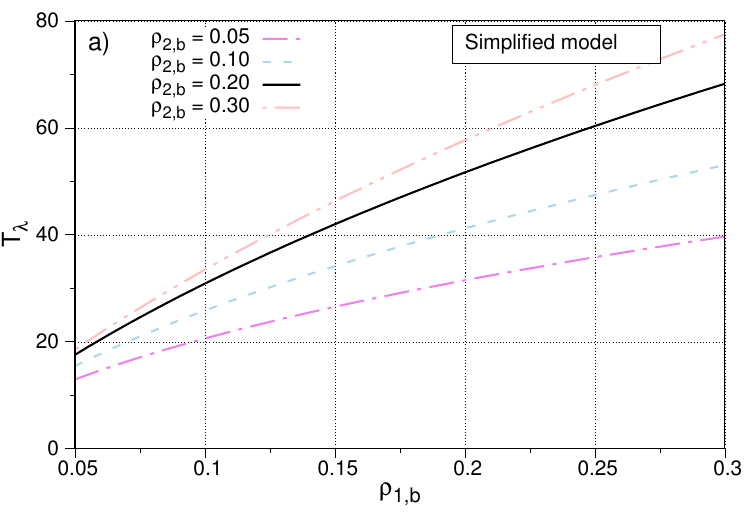}\hfill
    \includegraphics[width=0.49\textwidth]{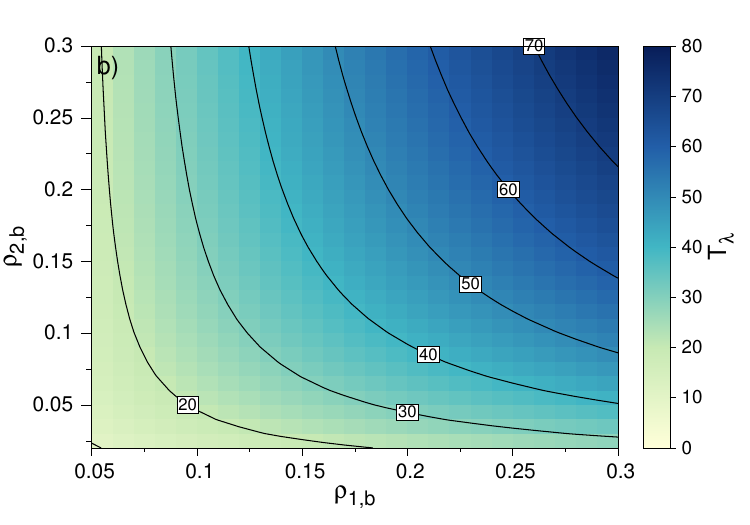}
    \caption{The $\lambda$-surface of the simplified model: the temperature
    $T_{\lambda}$ below which the bulk mixture ceases to be mesoscopically homogeneous.
    (a)~$\lambda$-lines, $T_{\lambda}$ as a function of $\rho_{1b}$ at fixed solvent
    densities $\rho_{2b}=0.05,0.10,0.20,0.30$; (b)~$T_{\lambda}$ over the
    $(\rho_{1b},\rho_{2b})$ plane (colour scale), with contour lines spaced by $10$.
    The linearized theory of this section applies above this surface.}
    \label{fig:lamline}
}
\end{center}
\end{figure}

The linearized description presupposes that the bulk mixture is mesoscopically
homogeneous, and it is worth recording where that is the case. The boundary of the
homogeneous region is the $\lambda$-surface -- the counterpart, for a
two-component mixture, of the $\lambda$-line of a one-component fluid -- reached
when a complex-conjugate pair
$\lambda\pm\ri\mu$ attains $\lambda=0$, so that the damping of the
oscillations disappears while their wavelength stays finite. Equivalently, it is
the locus at which $\min_{q}\det[\mathbf{I}+\beta\rho_{jb}\hat{\nu}_{ij}(q)]$ first
vanishes, at a non-zero wavenumber $q^{*}$. Figure~\ref{fig:lamline} shows the
resulting $T_{\lambda}(\rho_{1b},\rho_{2b})$ for the simplified model. 
Since the determinant depends on the densities and the temperature only through
the products $\beta\rho_{ib}$, $T_{\lambda}$ is a homogeneous function of degree
one of the two densities, $T_\lambda(s\rho_{1b},s\rho_{2b})=s\,T_\lambda(\rho_{1b},\rho_{2b})$,
and $q^{*}$ depends only on their ratio $\rho_{2b}/\rho_{1b}$.

Two features are relevant to what follows. First, within the present soft-core mean-field treatment $\hat\nu_{11}(q)>0$ and the pure fluid shows no $\lambda$-line. Excluded-volume effects, not included here, make it microsegregate at lower temperatures~\cite{Kravtsiv2022}. The instability of the mixture is therefore produced by the coupling to the solvent rather than by the competing interaction alone. Second, because the
cross and solvent interactions of the simplified model are purely repulsive, and
both strong and long-ranged, $T_{\lambda}$ rises steeply with either density. The
analytical theory is accordingly useful at relatively low densities for a given
temperature, and all the states considered below lie above the surface of
figure~\ref{fig:lamline}.

\vspace{2cm}
\subsection{Boundary conditions at the wall}
\label{sec:bc}

The eigenvalue problem of section~\ref{sec:lin} fixes the shapes
$\re^{-\lambda_{0}z}$ and $\re^{-\lambda z\pm\ri\mu z}$ of the admissible
solutions, but not their amplitudes and these must be 
supplemented with boundary conditions at the wall. Such conditions follow from the geometry alone. For $z<0$ both densities
vanish, so each potential obeys $V_{i}''=\alpha_{i}^{2}V_{i}$ there and, being
bounded as $z\to-\infty$, decays as $V_{i}(z)=V_{i}(0)\re^{\alpha_{i}z}$.
Continuity of $V_{i}$ and of $E_{i}=-V_{i}'$ at $z=0$ eliminates the single free
constant of this exterior solution and leaves one relation per field,
\begin{align}
\label{bcfield}
    V_{i}'(0)=\alpha_{i}V_{i}(0),\qquad\textrm{equivalently}\qquad
    E_{i}(0)=-\alpha_{i}V_{i}(0),\qquad i=1,2,3.
\end{align}
Equations~(\ref{bcfield}) are exact: nothing has been assumed about the density
profile, only that there is no fluid behind the wall and that the potential of a
bounded density distribution is continuously differentiable. They are in fact the
same statement that yields the contact theorem in section~\ref{sec:profiles}, at
the step where $\alpha_{i}^{2}V_{i}^{2}-E_{i}^{2}$ is seen to vanish at
$z=0_{+}$. Since there is one such relation for every Yukawa field, the number of
conditions equals the number of unknown amplitudes and nothing further is required.
In particular the contact densities are not prescribed: they follow from the
solution.

Evaluating $V_{i}(0)$ and $E_{i}(0)$ from the general solution (see
appendix~\ref{app:corrbc}), equations~(\ref{bcfield}) become a linear system for
the amplitudes,
\begin{align}
\label{bcnewsys}
    (\lambda_{0}+\alpha_{i})\bar{r}_{i0}
    +\big(\lambda^{2}+\mu^{2}+\alpha_{i}\lambda\big)\bar{r}_{i1}
    -\alpha_{i}\mu\,\bar{r}_{i2}=-\frac{\varkappa_{i}^{2}}{\alpha_{i}},
    \qquad i=1,2,3,
\end{align}
where the index $i$ labels the Yukawa term and not the species: $\alpha_{i}$ is
the range of the $i$th term of $\nu_{11}$, and the corresponding couplings are
$\varkappa_{1}^{2}\equiv\varkappa_{11}^{2}+\varkappa_{\mathrm s}^{2}$,
$\varkappa_{2}^{2}\equiv\varkappa_{12}^{2}$ and
$\varkappa_{3}^{2}\equiv\varkappa_{13}^{2}$. As in the eigenvalue problem, the
two-component character enters only through the solvent coupling
$\varkappa_{\mathrm s}^{2}$ contained in $\varkappa_{1}^{2}$. The elimination of
$\bar{r}_{20}$, $\bar{r}_{30}$ and $\bar{r}_{lm}$ and the resulting closed-form
amplitudes are given in appendix~\ref{app:corrbc}.

It is worth contrasting~(\ref{bcfield}) with the prescription used in the earlier
treatments of one-, two- and three-Yukawa fluids at a hard wall~\cite{HolovkoKravtsivSoviak2009, DiCaprio2011, Kravtsiv2015, Kravtsiv2022}, 
where the contact theorem is imposed on each of the three potentials separately,
\begin{align}
\label{bcold}
    V_{ib}-V_{i}(0)=\frac{\varkappa_{i}^{2}}{2\alpha_{i}^{2}},\qquad i=1,2,3 .
\end{align}
Summed over $i$, equations~(\ref{bcold}) reproduce the contact theorem exactly.
Taken separately they say more than it does: each states that a given Yukawa
potential at the wall is one half of its bulk value, which is what one obtains if
the fluid retains its bulk density right up to the surface. The contact densities
are then fixed at $\beta P_{1}^{\textrm{MF}}$ and $\beta P_{2}^{\textrm{MF}}$, a
species-resolved split that the exact nonlinear solution does not reproduce. The
two prescriptions differ at second order in the interaction: for a one-component
fluid, (\ref{bcold}) gives $\rho(0_{+})/\rho_{b}=1+\Gamma/2$ with
$\Gamma=\sum_{i}V_{ib}$, whereas~(\ref{bcfield}) yields
$\rho(0_{+})/\rho_{b}=\sqrt{1+\Gamma}$. The two agree to first order, which is the
order to which the linearization~(\ref{lin}) is valid, but $1+\Gamma/2$ is a value
imported from the nonlinear theory and imposed on the linearized equations, while
$\sqrt{1+\Gamma}$ is what those equations produce on their own.
Neither prescription is free of objection, and in what follows we examine both,
comparing each with the numerical solution of the nonlinear equations. It should
be borne in mind that a linearized theory cannot satisfy the contact theorem
identically in any case: the relation~(\ref{contact}) follows from the first
integral of the \emph{nonlinear} equations, and the truncation~(\ref{lin}) has
already discarded the terms needed to reproduce it. Retaining higher orders in the
expansion of the exponentials would bring the linearized contact densities closer
to it. Conditions~(\ref{bcold}) thus restore the summed relation by construction, but
at the price of a further assumption, namely that the contact density is split
between the species according to their partial pressures. This assumption is
supported neither by the contact theorem nor by the exact solution.

\subsection{Density profiles}
\label{sec:an_profiles}

Integrating the fields and substituting into the linearized
equations~(\ref{lin}), the density profiles of the two species take the form
\begin{align}
\label{profile1}
    \frac{\rho_{1}(z)}{\rho_{1b}}&=1-\big[\bar{r}_{10}+\bar{r}_{20}+\bar{r}_{30}\big]\re^{-\lambda_{0}z}
    -\re^{-\lambda z}\Big[\big\{\lambda(\bar{r}_{11}+\bar{r}_{21}+\bar{r}_{31})
    -\mu(\bar{r}_{12}+\bar{r}_{22}+\bar{r}_{32})\big\}\cos\mu z\nonumber\\
    &\qquad-\big\{\mu(\bar{r}_{11}+\bar{r}_{21}+\bar{r}_{31})
    +\lambda(\bar{r}_{12}+\bar{r}_{22}+\bar{r}_{32})\big\}\sin\mu z\Big],\\
\label{profile2}
    \frac{\rho_{2}(z)}{\rho_{2b}}&=1-\bar{r}_{10}\re^{-\lambda_{0}z}
    -\re^{-\lambda z}\Big[\big\{\lambda\bar{r}_{11}-\mu\bar{r}_{12}\big\}\cos\mu z
    -\big\{\mu\bar{r}_{11}+\lambda\bar{r}_{12}\big\}\sin\mu z\Big],
\end{align}
where $\bar{r}_{i0}$, $\bar{r}_{i1}$, $\bar{r}_{i2}$ ($i=1,2,3$) are the amplitudes
of the three mean-field potentials $V_{i}(z)$ and they are lengthy but explicit
functions of the pair-potential parameters, the densities and the temperature,
fixed by the eigenvector relations of $\mathbf{A}$ together with the boundary
conditions (\ref{bcfield}) of section~\ref{sec:bc}. Their closed-form
expressions are derived in appendix~\ref{app:corrbc}.
A notable feature of equations~(\ref{profile1})--(\ref{profile2}) is that the two species share the same decay coefficients $\lambda_{0}$, $\lambda$, and $\mu$: the profile of the main
component collects all three Yukawa fields, while that of the solvent involves only
the field with the range $\alpha_{1}$ common to both species. Each profile combines
a monotonic contribution that decays with the decay coefficient $\lambda_{0}$ with an oscillatory
contribution that decays with the decay coefficient $\lambda$ and oscillates with the wavelength
$2\piup/\mu$. This shared oscillatory structure, with amplitudes of opposite sign
for the two species, is precisely what produces their alternating
layering.

\subsection{Comparison with the numerical results}
\label{sec:an_compare}

Throughout this comparison the bulk densities are small enough
for the near-wall modulation to remain weak, which is the regime in which the
linearization~(\ref{lin}) is meaningful. Where the profile is strongly
structured, the linearization itself fails, and neither prescription for the
amplitudes is reliable.

\begin{figure}[htb]
\begin{center}
{
    \includegraphics[width=0.49\textwidth]{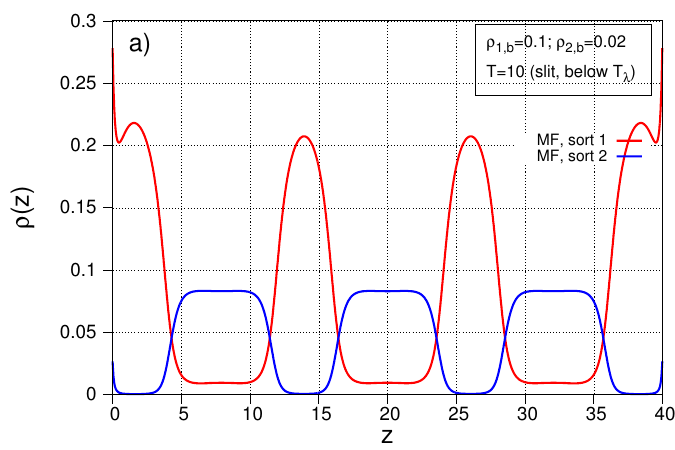}\hfill
    \includegraphics[width=0.49\textwidth]{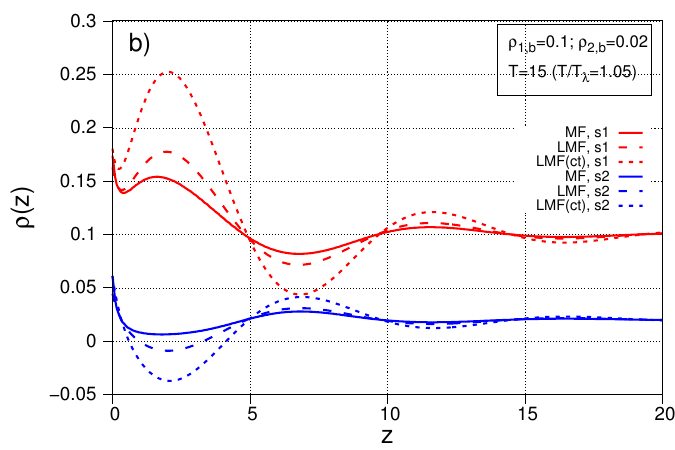}\\[1ex]
    \includegraphics[width=0.49\textwidth]{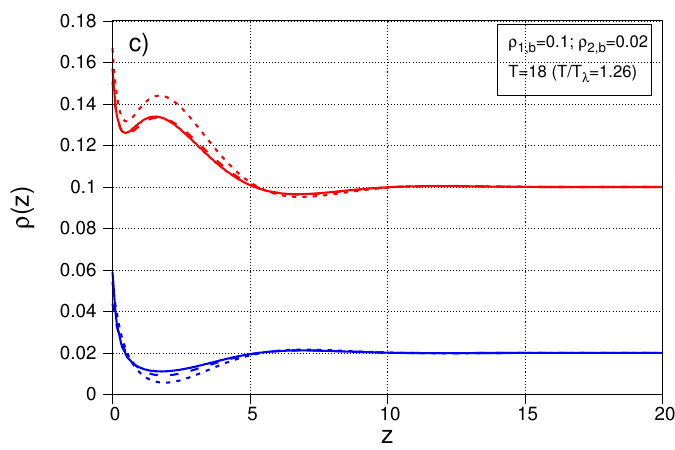}\hfill
    \includegraphics[width=0.49\textwidth]{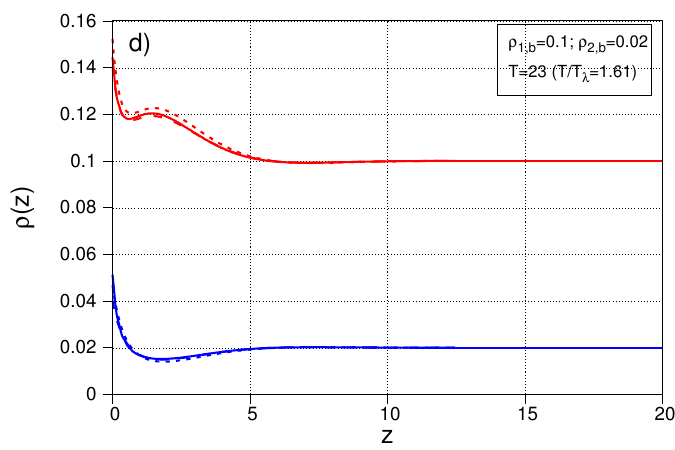}
    \caption{Effect of the boundary conditions on the density profiles at
    $\rho_{1b}=0.1$, $\rho_{2b}=0.02$ ($T_{\lambda}=14.3$): (a)~$T=10$, (b)~$T=15$,
    (c)~$T=18$, (d)~$T=23$. In (a), below $T_{\lambda}$, the mean-field solution in a
    slit of width $L_{z}=40$ develops alternating layers throughout and the linearized
    description does not apply, so only the MF profiles are shown. In (b)-(d), at a
    single hard wall: solid lines, numerical mean-field solution (MF); dashed lines,
    linearized profiles from the continuous conditions~(\ref{bcfield}); dotted lines,
    from the contact-value conditions~(\ref{bcold}) imposed on each potential
    separately. Red lines refer to the main component (sort~1), blue lines to the
    solvent (sort~2).}
\label{fig:bc_compare}
}
\end{center}
\end{figure}

The role of the boundary conditions is shown directly in
figure~\ref{fig:bc_compare}. Close to the boundary of the mesoscopically
homogeneous region the two prescriptions differ substantially. 
At $T=15$, i.e., $T/T_{\lambda}=1.05$, the first maximum of the main component
lies $0.054$ above the bulk density in the numerical solution, $0.077$ above it
with the continuity conditions~(\ref{bcfield}), and $0.152$ above it, nearly
three times the numerical value, with the contact-value
conditions~(\ref{bcold}). Over $0.5<z<25$, the root-mean-square deviation from
the numerical profiles obtained with~(\ref{bcfield}) is smaller than that
obtained with~(\ref{bcold}) by a factor of four for the main component and of
three for the solvent. Both linearized solutions take
slightly negative solvent densities near $z\approx2$ at this temperature, $-0.009$
with~(\ref{bcfield}) against $-0.037$ with~(\ref{bcold}), which shows that the
linearization itself is close to its limit so near $T_{\lambda}$. The difference
between the two prescriptions falls off rapidly with temperature: at $T=18$ the
corresponding factors are eight and three, and by $T=23$ the three sets of curves
are indistinguishable on the scale of the figure~\ref{fig:bc_compare}, panels (a)--(d).

The two prescriptions differ correspondingly in their behaviour at contact.
Equations~(\ref{bcold}) reproduce the summed contact theorem~(\ref{contact}) by
construction, at the cost of the profile away from the wall. With the exact
conditions~(\ref{bcfield}) the contact densities are outputs rather than inputs,
and their sum falls below $\beta P^{\textrm{MF}}$ by $14$, $12$ and $9$ percent at
$T=15$, $18$ and $23$ respectively. This is the second-order discrepancy discussed in section~\ref{sec:bc}. 
It diminishes as the temperature rises and the linearization~(\ref{lin}) becomes
more accurate. In neither case are the species-resolved relations $\beta P_{i}^{\textrm{MF}}=\rho_{i}(0_{+})$ obeyed, and the numerical solution does not obey them either: only the summed
relation~(\ref{contact}) is exact.

\subsection{Qualitative behaviour with temperature and solvent density}
\label{sec:an_qual}

The analytical expressions~(\ref{profile1})--(\ref{profile2}) provide a convenient
tool to explore the qualitative behaviour of the profiles as functions of the
thermodynamic parameters. {As the temperature is varied at fixed densities $\rho_{1b}=0.1$,
$\rho_{2b}=0.02$ (figure~\ref{fig:bc_compare}), on cooling towards $T_\lambda$ the
amplitude of the near-surface structure grows and the decay length $1/\lambda$
increases, so that the alternating oscillations of the two species extend
further into the bulk. On heating, the structure becomes weak and the mixture
is essentially uniform already a few diameters away from the wall.}

\begin{figure}[htb]
\begin{center}
{
    \includegraphics[width=0.49\textwidth]{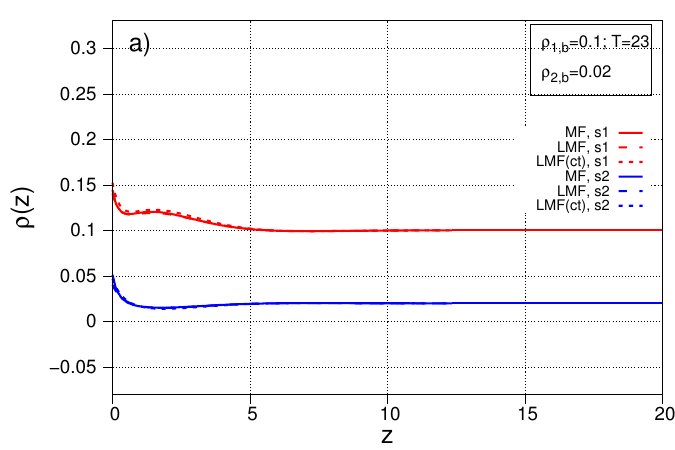}
    \includegraphics[width=0.49\textwidth]{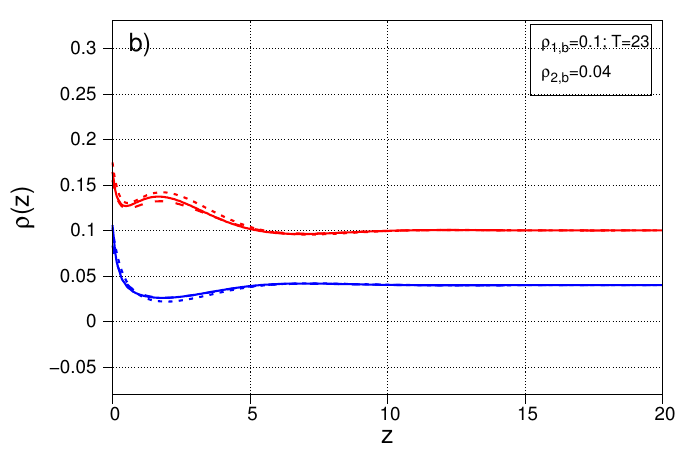} \\
    \includegraphics[width=0.49\textwidth]{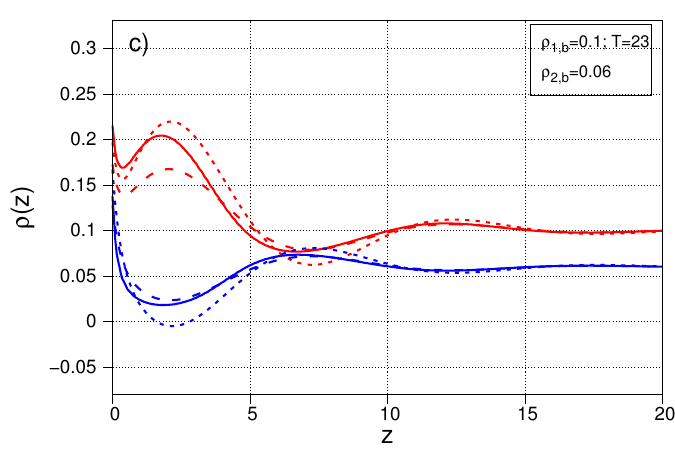}
    \includegraphics[width=0.49\textwidth]{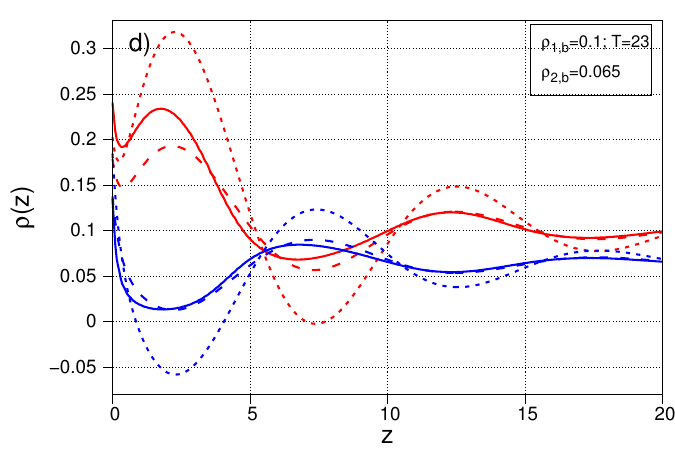}
\mbox{}
    \caption{Density profiles for $\rho_{1b}=0.1$ at $T=23$ and increasing solvent
    density: (a)~$\rho_{2b}=0.02$, (b)~$\rho_{2b}=0.04$,
    (c)~$\rho_{2b}=0.06$ and (d)~$\rho_{2b}=0.065$} (just above the $\lambda$-line). In each panel
    the numerical mean-field solution (MF, solid) is compared with the linearized
    profiles from the continuous conditions~(\ref{bcfield}) (dashed) and from the
    contact-value conditions~(\ref{bcold}) (dotted). Red lines refer to species~1
    (sort~1), blue lines to species~2 (sort~2).
    \label{fig:varRho}
}
\end{center}
\end{figure}

The role of the solvent concentration is illustrated in figure~\ref{fig:varRho},
where the density of the main component is fixed at $\rho_{1b}=0.1$ and
$T=23$ while the solvent density is increased from
$\rho_{2b}=0.02$ up to $\rho_{2b}=0.065$ (the last value lying just above the
$\lambda$-line) in four steps (all states still above the
$\lambda$-surface of figure~\ref{fig:lamline}, so that the linearization applies).
{At the lowest solvent density shown ($\rho_{2b}=0.02$) the
near-surface structure is weak and the three descriptions nearly coincide.} As the
solvent density grows, the
structure of both components becomes progressively more pronounced: the profile of
the main component develops a marked maximum near the surface, and the two species
arrange themselves in an increasingly clear microsegregated pattern. This
shows that the solvent, although interacting through a simple potential, actively
promotes the near-surface ordering and the microphase separation of the
mixture. Each panel of figure~\ref{fig:varRho} also compares the
two boundary-condition prescriptions with the numerical mean-field solution: at low
$\rho_{2b}$ all three coincide, but as $\rho_{2b}$ grows and the state approaches the
$\lambda$-line the contact-value prescription overshoots the near-wall maximum. 
For the state closest to the $\lambda$-surface, $\rho_{2b}=0.065$
(figure~\ref{fig:bc_compare}d), it develops large oscillations and negative
solvent densities, whereas the continuity conditions continue to track the
mean-field solution. This is the same trend seen with temperature in figure~\ref{fig:bc_compare}.

\section{Conclusions}
\label{sec:concl}

We have developed a field-theoretical mean-field description of a binary
mixture of soft particles near a planar surface, in which the main component interacts 
through a competing SALR potential (a sum of three Yukawa terms), while the solvent 
and the cross interactions are modelled by two Yukawa terms. We derived the coupled equations 
for the density profiles of both species, obtained a first integral of these equations, 
and used it to formulate the contact theorem for the two-component system, 
which reduces to the known one-component relation in the appropriate limit.

The mean-field equations were solved numerically in the bulk and for a fluid confined by hard walls. 
At the mean-field level, in the temperature range considered here, the pure SALR fluid is homogeneous -- it develops its own mesostructures only at lower temperatures~\cite{Kravtsiv2022}, so the solvent shifts the microsegregation onset up into the region where the one-component fluid is still uniform.
At high temperatures the two species are homogeneously mixed, whereas upon cooling or upon increasing 
the density they separate into alternating layers rich in one or the other component, 
the onset temperature depending on the composition.
Near a hard wall the main component is strongly adsorbed and retains the two-peak structure typical of the one-component fluid~\cite{Kravtsiv2022}, while the solvent is expelled from a near-surface layer a few particle diameters wide. The wall promotes the ordering, so that the layering persists across a slit up to temperatures at which the bulk mixture is already homogeneous. Molecular dynamics simulations confirm this picture. The mean-field theory overestimates the onset temperature of microsegregation by roughly a factor of two, but oncethis shift is taken into account the mean-field profiles agree well with the simulation data, 
particularly for the main component and for the denser mixture.

In the mesoscopically homogeneous region, above the $\lambda$-surface, we obtained closed-form density profiles by linearizing the theory and solving the resulting eigenvalue problem. When the solvent--solvent and cross interactions are purely repulsive and coincide in range and amplitude with the
leading Yukawa term of the main component, the problem reduces to a cubic equation with a single real root and a complex-conjugate pair, giving a monotonic plus an exponentially damped oscillatory contribution. Both species share the same decay lengths and oscillation period, with oscillatory
amplitudes of opposite sign, which is the origin of their alternating layering. The analytical profiles correctly describe the dependence on temperature and solvent concentration, in particular the solvent-induced
enhancement of the near-surface ordering. 
The amplitudes of these profiles are fixed by boundary conditions that follow from the continuity of the Yukawa potentials and of their derivatives at the wall. These conditions are exact and supply one relation per Yukawa field, so that their number matches the number of unknown amplitudes. The same
continuity underlies the contact theorem itself, which, however, yields only one summed relation. Imposing it on each potential separately, as done previously, is therefore an additional assumption, which in a mixture also prescribes an unjustified split of the contact density between the species. The continuity conditions require no such assumption and leave the contact densities to be determined by the solution. With these conditions the linearized profiles agree quantitatively with the numerical mean-field solution except very close to the wall, the agreement deteriorating only in the immediate vicinity of the $\lambda$-surface.

These results extend the field-theoretical description of one-component Yukawa fluids~\cite{Kravtsiv2015,Kravtsiv2022}, 
including the competing-interaction (SALR) case~\cite{Kravtsiv2022}, to two-component mixtures, and complement recent studies of self-assembling mixtures near surfaces and in
monolayers~\cite{PatsahanMeyraCiach2022,PatsahanMeyraCiach2024}. Natural extensions of the present work include the linearized treatment of the full $3\text{Y}{+}2\text{Y}$ model at hard and attractive walls, for which the seventh-degree characteristic equation must be solved numerically.

\section*{Acknowledgements}

IK and TP acknowledge financial support from the National Research Foundation
of Ukraine (project No.~2020.02/0317). OP acknowledges the support of the
Ministry of Education and Science of Ukraine (Agreement No.~PH/16-2023).
Computer time for the reported simulations was provided by the
Interdisciplinary Center for Computer Simulations (Lviv), which is supported by
the NRFU Grant No.~2023.05/0019.

\appendix

\section{Derivation of the linearized density profiles}
\label{app:derivation}

Here we give the details of the calculation summarized in
section~\ref{sec:analytical}. 
The scheme is a two-component generalization of the one-component derivation
of~\cite{Kravtsiv2022}, and is carried out for the simplified
model~(\ref{eq:simplified}), in which the solvent and cross interactions are
described by a single Yukawa term identical to the first term of $\nu_{11}$,
i.e., with the same amplitude $A_{11}^{(1)}$ and range $\alpha_{1}$.
Throughout we use the abbreviations of section~\ref{sec:lin}, i.e., $\alpha_{k}$ and
$\varkappa_{1k}^{2}$ ($k=1,2,3$) for the ranges and coupling parameters of the three
Yukawa terms of the main component, and $\varkappa_{\mathrm s}^{2}$ for the contribution of
the solvent through the shared term. Compared with the one-component derivation
of~\cite{Kravtsiv2022}, the two-component character enters in three places: the
solvent coupling $\varkappa_{\mathrm s}^{2}$, which appears in the $(1,1)$ entry of
$\mathbf{A}$ in equation~(\ref{matrix3}) and in the first boundary condition below;
the field superposition $\tilde{E}=E_{11}^{(1)}+E_{12}^{(1)}$ that closes the
reduced system; and the species-resolved profiles~(\ref{profile1})-(\ref{profile2}),
in which the main component collects all three Yukawa fields while the solvent
couples only to the shared one. The remaining algebra parallels the one-component
case, and is quoted compactly below.

\subsection{Roots of the characteristic cubic}
\label{app:roots}

The eigenvalues of the matrix $\mathbf{A}$ in equation~(\ref{matrix3}), written as
$\Lambda_{0}=\lambda_{0}^{2}$ and $\Lambda_{1,2}=\lambda_{1,2}^{2}$, are the roots
of the cubic~(\ref{cubic}) with the coefficients~(\ref{coefs}). Introducing
    \begin{align}
    \label{app:Delta}
    \Delta_{0}=b^{2}-3c,\qquad
    \Delta_{1}=2b^{3}-9bc+27d,\qquad
    C=\left[\frac{\Delta_{1}+\sqrt{\Delta_{1}^{2}-4\Delta_{0}^{3}}}{2}\right]^{1/3},
\end{align}
where $C$ is taken to be the real positive cube root, the real eigenvalue and the
complex-conjugate pair follow from the standard (Cardano) solution as
\begin{align}
\label{app:l0}
    \lambda_{0}^{2}&=\frac{1}{3}\left(b+C+\frac{\Delta_{0}}{C}\right),\qquad
    \lambda_{1,2}^{2}=M_{1}\pm\ri M_{2},\\
    \label{app:M}
    M_{1}&=\frac{1}{3}\left(b-\frac{C}{2}-\frac{\Delta_{0}}{2C}\right),\qquad
    M_{2}=\frac{\sqrt{3}}{6}\left(C-\frac{\Delta_{0}}{C}\right).
\end{align}
For the parameter range considered here $\lambda_{0}^{2}$ is real and positive,
while $\lambda_{1,2}^{2}$ are complex conjugate. Writing
$\lambda_{1,2}=\lambda\pm\ri\mu$ one obtains
\begin{align}
\label{app:lmu}
    \lambda_{0}=\sqrt{\lambda_{0}^{2}},\qquad
    \lambda=\frac{1}{\sqrt{2}}\Big[\sqrt{M_{1}^{2}+M_{2}^{2}}+M_{1}\Big]^{1/2},\qquad
    \mu=\frac{\operatorname{sgn}(M_{2})}{\sqrt{2}}\Big[\sqrt{M_{1}^{2}+M_{2}^{2}}-M_{1}\Big]^{1/2},
\end{align}
so that $\lambda_{0}$ and $\lambda$ are the (positive) 
decay coefficients of monotonic and oscillatory decay, respectively, and $2\piup/\mu$ is the oscillation wavelength.

\subsection{Field amplitudes}
\label{app:fields}

Besides the superposition $\tilde{E}(z)\equiv E_{11}^{(1)}(z)+E_{12}^{(1)}(z)$, the
two remaining fields associated with the second and third Yukawa terms of the main
component are denoted $E_{12}(z)$ and $E_{13}(z)$. Retaining only the contributions
that decay as $z\to\infty$, their general form is
\begin{align}
\label{app:Efields}
    \tilde{E}(z)&=c_{10}\re^{-\lambda_{0}z}+(r_{11}+\ri r_{12})\re^{(-\lambda+\ri\mu)z}+(r_{11}-\ri r_{12})\re^{(-\lambda-\ri\mu)z},\nonumber\\
    E_{12}(z)&=c_{20}\re^{-\lambda_{0}z}+(r_{21}+\ri r_{22})\re^{(-\lambda+\ri\mu)z}+(r_{21}-\ri r_{22})\re^{(-\lambda-\ri\mu)z},\nonumber\\
    E_{13}(z)&=c_{30}\re^{-\lambda_{0}z}+(r_{31}+\ri r_{32})\re^{(-\lambda+\ri\mu)z}+(r_{31}-\ri r_{32})\re^{(-\lambda-\ri\mu)z}.
\end{align}
Because equation~(\ref{matrix})-(\ref{matrix3}) is homogeneous, only three of these
amplitudes are independent, and we take $c_{10}$, $r_{11}$, $r_{12}$. Substituting
(\ref{app:Efields}) into $\mathbf{E}''=\mathbf{A}\mathbf{E}$ and equating the
coefficients of the same $z$-dependent function separately in their real and
imaginary parts gives two decoupled linear systems. 
For the oscillatory terms, with
$\bar\kappa_1^2=\lambda^2-\mu^2-\alpha_1^2-\varkappa_{11}^2-{\varkappa_{\mathrm s}^{2}}$ and
$\bar\kappa_j^2=\lambda^2-\mu^2-\alpha_j^2-\varkappa_{1j}^2$ ($j=2,3$),
\begin{align}
\label{app:osc}
    \bar{\kappa}_{1}^{2}r_{11}+2\lambda\mu\,r_{12}-\varkappa_{11}^{2}(r_{21}+r_{31})&=0,&
    \bar{\kappa}_{1}^{2}r_{12}-2\lambda\mu\,r_{11}-\varkappa_{11}^{2}(r_{22}+r_{32})&=0,\nonumber\\
    \bar{\kappa}_{2}^{2}r_{21}+2\lambda\mu\,r_{22}-\varkappa_{12}^{2}(r_{11}+r_{31})&=0,&
    \bar{\kappa}_{2}^{2}r_{22}-2\lambda\mu\,r_{21}-\varkappa_{12}^{2}(r_{12}+r_{32})&=0,\nonumber\\
    \bar{\kappa}_{3}^{2}r_{31}+2\lambda\mu\,r_{32}-\varkappa_{13}^{2}(r_{11}+r_{21})&=0,&
    \bar{\kappa}_{3}^{2}r_{32}-2\lambda\mu\,r_{31}-\varkappa_{13}^{2}(r_{12}+r_{22})&=0.
\end{align}
Any two of these pairs suffice. Solving them expresses the remaining oscillatory
amplitudes linearly through $r_{11}$ and $r_{12}$,
$r_{lm}=K_{lm}^{11}r_{11}+K_{lm}^{12}r_{12}$ ($l=2,3$), while matching the monotonic
terms gives $c_{20}=K_{20}c_{10}$ and $c_{30}=K_{30}c_{10}$. The coefficients
$K_{lm}^{ij}$, $K_{20}$ and $K_{30}$ are rational functions of
$\{\alpha_{k},\varkappa_{1k}^{2},\lambda,\mu\}$; their explicit form is obtained
exactly as in the one-component case~\cite{Kravtsiv2022}.

\subsection{Potentials, boundary conditions and coefficients}
\label{app:bc}

Integrating the fields once yields the mean-field potentials
\begin{align}
\label{app:V}
    V_{i}(z)=V_{ib}+\bar{r}_{i0}\re^{-\lambda_{0}z}
    +\re^{-\lambda z}\big[(\lambda\bar{r}_{i1}-\mu\bar{r}_{i2})\cos\mu z
    -(\mu\bar{r}_{i1}+\lambda\bar{r}_{i2})\sin\mu z\big],
\end{align}
where $\bar{r}_{i0}=+c_{i0}/\lambda_{0}$ and
$\bar{r}_{ij}=2r_{ij}/(\lambda^{2}+\mu^{2})$ ($i=1,2,3$, $j=1,2$); both follow
from $V_{i}(z)=V_{ib}+\int_{z}^{\infty}E_{i}(t)\,\rd t$, which is the integral
of~(\ref{system1}) with $V_{i}\to V_{ib}$ as $z\to\infty$. 
The relations of section~\ref{app:fields} carry over to the barred amplitudes, 
so that $\bar{r}_{20}=K_{20}\bar{r}_{10}$,
$\bar{r}_{30}=K_{30}\bar{r}_{10}$ and
$\bar{r}_{lm}=K_{lm}^{11}\bar{r}_{11}+K_{lm}^{12}\bar{r}_{12}$. The three remaining
unknowns $\bar{r}_{10}$, $\bar{r}_{11}$, $\bar{r}_{12}$ are fixed by the boundary
conditions at the wall. The conditions~(\ref{bcfield}) used in this work are
treated in section~\ref{app:corrbc}. For comparison we first record the earlier
prescription, in which the contact theorem is imposed on each potential
separately. Setting $z=0$ in the linearized
relations gives $V_{ib}-V_{i}(0)=\varkappa_{i}^{2}/(2\alpha_{i}^{2})$, i.e.,
\begin{align}
\label{app:bcsys}
    -\frac{\varkappa_{11}^{2}+\varkappa_{\mathrm s}^{2}}{2\alpha_{1}^{2}}&=\bar{r}_{10}+\lambda\bar{r}_{11}-\mu\bar{r}_{12},\qquad
    -\frac{\varkappa_{12}^{2}}{2\alpha_{2}^{2}}=\bar{r}_{20}+\lambda\bar{r}_{21}-\mu\bar{r}_{22},\nonumber\\
    -\frac{\varkappa_{13}^{2}}{2\alpha_{3}^{2}}&=\bar{r}_{30}+\lambda\bar{r}_{31}-\mu\bar{r}_{32}.
\end{align}
Together with the amplitude relations of section~\ref{app:fields}, the three
equations~(\ref{app:bcsys}) form a closed linear system for $\bar{r}_{10}$,
$\bar{r}_{11}$, $\bar{r}_{12}$, from which these coefficients are obtained in closed
form. Note that the two-component nature of the problem is confined to the first of
equations~(\ref{app:bcsys}), through the term $\varkappa_{\mathrm s}^{2}$. 
The solution otherwise follows the same algebra as the one-component case~\cite{Kravtsiv2022}.
The amplitudes used for the profiles~(\ref{profile1})
and~(\ref{profile2}) of the main text are those of section~\ref{app:corrbc}, obtained
from the exact conditions~(\ref{bcfield}) instead.

\subsection{Amplitudes for the corrected boundary conditions}
\label{app:corrbc}

We collect here the algebra behind the conditions~(\ref{bcfield}) introduced in
section~\ref{sec:bc}. Setting $z=0$ in~(\ref{app:V}) gives
$V_{i}(0)-V_{ib}=\bar{r}_{i0}+\lambda\bar{r}_{i1}-\mu\bar{r}_{i2}$, which is what
produced~(\ref{app:bcsys}), while differentiating~(\ref{app:V}) at $z=0$ gives
\begin{align}
\label{app:E0}
    V_{i}'(0)=-\lambda_{0}\bar{r}_{i0}-(\lambda^{2}+\mu^{2})\bar{r}_{i1},
    \qquad\textrm{i.e.}\qquad
    E_{i}(0)=\lambda_{0}\bar{r}_{i0}+(\lambda^{2}+\mu^{2})\bar{r}_{i1} .
\end{align}
Substituting both into~(\ref{bcfield}) and using
$V_{1b}=(\varkappa_{11}^{2}+\varkappa_{\mathrm s}^{2})/\alpha_{1}^{2}$,
$V_{2b}=\varkappa_{12}^{2}/\alpha_{2}^{2}$ and
$V_{3b}=\varkappa_{13}^{2}/\alpha_{3}^{2}$ yields the system~(\ref{bcnewsys}) of
the main text, which replaces~(\ref{app:bcsys}).

Eliminating $\bar{r}_{20}$, $\bar{r}_{30}$ and $\bar{r}_{lm}$ by the amplitude
relations of section~\ref{app:fields} reduces~(\ref{bcnewsys}) to a closed
$3\times3$ system for the three remaining unknowns,
\begin{align}
\label{app:cram}
    a_{i}\bar{r}_{10}+b_{i}\bar{r}_{11}+c_{i}\bar{r}_{12}=d_{i},\qquad i=1,2,3,
\end{align}
with $W_{i}=\lambda^{2}+\mu^{2}+\alpha_{i}\lambda$ and
\begin{align}
    a_{i}&=(\lambda_{0}+\alpha_{i})K_{i0},&
    b_{i}&=W_{i}K_{i1}^{11}-\alpha_{i}\mu K_{i2}^{11},\nonumber\\
    c_{i}&=W_{i}K_{i1}^{12}-\alpha_{i}\mu K_{i2}^{12},&
    d_{i}&=-\frac{\varkappa_{i}^{2}}{\alpha_{i}},\label{app:abcd}
\end{align}
where the conventions $K_{10}=K_{11}^{11}=K_{12}^{12}=1$,
$K_{11}^{12}=K_{12}^{11}=0$ complete the set of amplitude coefficients of
section~\ref{app:fields}. For comparison, the earlier
conditions~(\ref{app:bcsys}) are the same system~(\ref{app:cram}) with
$a_{i}=K_{i0}$, $b_{i}=\lambda K_{i1}^{11}-\mu K_{i2}^{11}$,
$c_{i}=\lambda K_{i1}^{12}-\mu K_{i2}^{12}$ and
$d_{i}=-\varkappa_{i}^{2}/(2\alpha_{i}^{2})$.

Since~(\ref{app:cram}) is a $3\times3$ linear system with explicit coefficients,
Cramer's rule gives the amplitudes in closed form,
\begin{align}
\label{app:cramsol}
    \bar{r}_{10}=\frac{D_{1}}{D},\qquad
    \bar{r}_{11}=\frac{D_{2}}{D},\qquad
    \bar{r}_{12}=\frac{D_{3}}{D},
\end{align}
where $D$ is the determinant of the coefficient matrix of~(\ref{app:cram}) and
$D_{1}$, $D_{2}$, $D_{3}$ are obtained from it by replacing its first, second and
third column, respectively, by the column $(d_{1},d_{2},d_{3})^{\textrm{T}}$,
\begin{align}
\label{app:dets}
    D=\left|
    \begin{array}{ccc}
    a_{1}&b_{1}&c_{1}\\
    a_{2}&b_{2}&c_{2}\\
    a_{3}&b_{3}&c_{3}
\end{array}\right|,
\quad
    D_{1}=\left|
\begin{array}{ccc}
    d_{1}&b_{1}&c_{1}\\
    d_{2}&b_{2}&c_{2}\\
    d_{3}&b_{3}&c_{3}
\end{array}\right|,
\quad
D_{2}=\left|
\begin{array}{ccc}
    a_{1}&d_{1}&c_{1}\\
    a_{2}&d_{2}&c_{2}\\
    a_{3}&d_{3}&c_{3}
\end{array}\right|,
\quad
D_{3}=\left|
\begin{array}{ccc}
    a_{1}&b_{1}&d_{1}\\
    a_{2}&b_{2}&d_{2}\\
    a_{3}&b_{3}&d_{3}
\end{array}\right|.
\end{align}
The remaining amplitudes follow from the relations of section~\ref{app:fields}, and
the profiles from~(\ref{profile1})-(\ref{profile2}) as before. Because the
cubic~(\ref{cubic}) is itself solvable in radicals
through~(\ref{app:l0})-(\ref{app:lmu}), the entire solution remains explicit.
Setting $z=0$ in~(\ref{profile1})-(\ref{profile2}), the contact densities that
result are
\begin{align}
\label{app:contact}
    \frac{\rho_{1}(0_{+})}{\rho_{1b}}
    =1-\sum_{i=1}^{3}\big(\bar{r}_{i0}+\lambda\bar{r}_{i1}-\mu\bar{r}_{i2}\big),
    \qquad
    \frac{\rho_{2}(0_{+})}{\rho_{2b}}
    =1-\big(\bar{r}_{10}+\lambda\bar{r}_{11}-\mu\bar{r}_{12}\big),
\end{align}
neither of which is imposed.


\begin{thebibliography}{10}
	\providecommand{\url}[1]{\texttt{#1}}
	\providecommand{\urlprefix}{URL }
	\expandafter\ifx\csname urlstyle\endcsname\relax
	\providecommand{\doi}[1]{doi:\discretionary{}{}{}#1}\else
	\providecommand{\doi}{doi:\discretionary{}{}{}\begingroup
		\urlstyle{rm}\Url}\fi
	\providecommand{\eprint}[2][]{\url{#2}}
	
	\bibitem{Sweatman2019}
	Sweatman~M.~B., Lue~L., The giant {SALR} cluster fluid: A review, Adv. Theory
	Simul., 2019, \textbf{2}, 1900025, \doi{10.1002/adts.201900025}.
	
	\bibitem{vonKonigslow_2013}
	von Konigslow~K., Cardenas-Mendez~E.~D., Thompson~R.~B., Rasmussen~K.~{\O}.,
	The self-assembly of particles with isotropic interactions, J. Phys.:
	Condens. Matter, 2013, \textbf{25}, 325101,
	\doi{10.1088/0953-8984/25/32/325101}.
	
	\bibitem{FerreiroRangel2018}
	Ferreiro-Rangel~C.~A., Sweatman~M.~B., Cluster formation in binary fluids with
	competing short-range and long-range interactions, Mol. Phys., 2018,
	\textbf{116}, 3231--3244, \doi{10.1080/00268976.2018.1489082}.
	
	\bibitem{Tan2021}
	Tan~J., Afify~N.~D., Ferreiro-Rangel~C.~A., Fan~X., Sweatman~M.~B., Cluster
	formation in symmetric binary {SALR} mixtures, J. Chem. Phys., 2021,
	\textbf{154}, 074504, \doi{10.1063/5.0036046}.
	
	\bibitem{TanSweatman2024}
	Tan~J., Sweatman~M.~B., Secondary nucleation in symmetric binary {SALR}
	mixtures, Phys. Chem. Chem. Phys., 2024, \textbf{26}, 17057,
	\doi{10.1039/D3CP05765H}.
	
	\bibitem{Munao2022}
	Muna{\`o}~G., Prestipino~S., Bomont~J.-M., Costa~D., Clustering in mixtures of
	{SALR} particles and hard spheres with cross attraction, J. Phys. Chem. B,
	2022, \textbf{126}, 2027, \doi{10.1021/acs.jpcb.1c09758}.
	
	\bibitem{munao:22:0}
	Muna{\`o}~G., Costa~D., Malescio~G., Bomont~J.-M., Prestipino~S., Competition
	between clustering and phase separation in binary mixtures containing {SALR}
	particles, Soft Matter, 2022, \textbf{18}, 6453--6464,
	\doi{10.1039/d2sm00944g}.
	
	\bibitem{PatsahanLitniewskiCiach2021}
	Patsahan~O., Litniewski~M., Ciach~A., Self-assembly in mixtures with competing
	interactions, Soft Matter, 2021, \textbf{17}, 2883, \doi{10.1039/D0SM02072A}.
	
	\bibitem{PatsahanMeyraCiach2022}
	Patsahan~O., Meyra~A., Ciach~A., Effect of a confining surface on a mixture
	with spontaneous inhomogeneities, J. Mol. Liq., 2022, \textbf{363}, 119844,
	\doi{10.1016/j.molliq.2022.119844}.
	
	\bibitem{PatsahanMeyraCiach2024}
	Patsahan~O., Meyra~A., Ciach~A., Spontaneous pattern formation in monolayers of
	binary mixtures with competing interactions, Soft Matter, 2024, \textbf{20},
	1410, \doi{10.1039/D3SM01537H}.
	
	\bibitem{Prestipino2023}
	Prestipino~S., Pini~D., Costa~D., Malescio~G., Muna{\`o}~G., A density
	functional theory and simulation study of stripe phases in symmetric
	colloidal mixtures, J. Chem. Phys., 2023, \textbf{159}, No.~20, 204902,
	\doi{10.1063/5.0177209}.
	
	\bibitem{Pini2025}
	Pini~D., Prestipino~S., Costa~D., Malescio~G., Muna{\`o}~G., A
	density-functional theory study of vapor, liquid, and mesophases in a
	symmetric, hard-sphere mixture with cross attraction at non-equimolar
	concentrations, J. Chem. Phys., 2025, \textbf{163}, No.~16, 164905,
	\doi{10.1063/5.0291013}.
	
	\bibitem{sweatman2019assembly}
	Sweatman~M.~B., Insall~R., Assembly of the actin catalyst {WASP} by giant
	{SALR} cluster formation, Advanced Theory and Simulations, 2019, \textbf{2},
	1800203, \doi{10.1002/adts.201800203}.
	
	\bibitem{ArcherWilding2007}
	Archer~A.~J., Wilding~N.~B., Phase behavior of a fluid with competing
	attractive and repulsive interactions, Phys. Rev. E, 2007, \textbf{76},
	031501, \doi{10.1103/PhysRevE.76.031501}.
	
	\bibitem{Ciach2008}
	Ciach~A., Universal sequence of ordered structures obtained from mesoscopic
	description of self-assembly, Phys. Rev. E, 2008, \textbf{78}, No.~6, 061505,
	\doi{10.1103/physreve.78.061505}.
	
	\bibitem{Ciach2010}
	Ciach~A., G{\'o}{\'z}d{\'z}~W., Mesoscopic description of network-forming
	clusters of weakly charged colloids, Condens. Matter Phys., 2010,
	\textbf{13}, No.~2, 23603, \doi{10.5488/cmp.13.23603}.
	
	\bibitem{Ciach2013}
	Ciach~A., P{\c e}kalski~J., G\'o\'zd\'z~W.~T., Origin of similarity of phase
	diagrams in amphiphilic and colloidal systems with competing interactions,
	Soft Matter, 2013, \textbf{9}, No.~27, 6301, \doi{10.1039/c3sm50668a}.
	
	\bibitem{zhuang:16:0}
	Zhuang~Y., Zhang~K., Charbonneau~P., Equilibrium phase behavior of a
	continuous-space microphase former, Phys. Rev. Lett., 2016, \textbf{116},
	098301, \doi{10.1103/PhysRevLett.116.098301}.
	
	\bibitem{Edelmann2016}
	Edelmann~M., Roth~R., Gyroid phase of fluids with spherically symmetric
	competing interactions, Phys. Rev. E, 2016, \textbf{93}, No.~6, 062146,
	\doi{10.1103/physreve.93.062146}.
	
	\bibitem{pini:17:0}
	Pini~D., Parola~A., Pattern formation and self-assembly driven by competing
	interactions, Soft Matter, 2017, \textbf{13}, 9259, \doi{10.1039/C7SM02125A}.
	
	\bibitem{ZhuangCharbonneau2016}
	Zhuang~Y., Charbonneau~P., Recent advances in the theory and simulation of
	model colloidal microphase formers, J. Phys. Chem. B, 2016, \textbf{120},
	7775, \doi{10.1021/acs.jpcb.6b05471}.
	
	\bibitem{Bretonnet2019}
	Bretonnet~J.-L., Competing interactions in colloidal suspensions, AIMS Mater.
	Sci., 2019, \textbf{6}, 509, \doi{10.3934/matersci.2019.4.509}.
	
	\bibitem{costa:23:0}
	Costa~D., Muna{\`o}~G., Bomont~J.-M., Malescio~G., Palatella~A., Prestipino~S.,
	Microphase versus macrophase separation in the square-well-linear fluid: A
	theoretical and computational study, Phys. Rev. E, 2023, \textbf{108},
	034602, \doi{10.1103/PhysRevE.108.034602}.
	
	\bibitem{Serna2026}
	Serna~H., Mart{\'\i}n-Roca~J., Meyra~A.~G., Noya~E.~G., Activity enhances
	transport while competing interactions preserve structure in colloidal
	microphase formers, J. Chem. Phys., 2026, \textbf{165}, 054902,
	\doi{10.1063/5.0343447}.
	
	\bibitem{Leunissen2005}
	Leunissen~M.~E., Christova~C.~G., Hynninen~A.-P., Royall~C.~P., Campbell~A.~I.,
	Imhof~A., Dijkstra~M., van Roij~R., van Blaaderen~A., Ionic colloidal
	crystals of oppositely charged particles, Nature, 2005, \textbf{437}, 235,
	\doi{10.1038/nature03946}.
	
	\bibitem{Campbell2005}
	Campbell~A.~I., Anderson~V.~J., van Duijneveldt~J.~S., Bartlett~P., Dynamical
	arrest in attractive colloids: The effect of long-range repulsion, Phys. Rev.
	Lett., 2005, \textbf{94}, 208301, \doi{10.1103/PhysRevLett.94.208301}.
	
	\bibitem{Dhumal2026}
	Dhumal~U., Cross-interaction softness as a route to microphase separation in
	binary colloidal systems, 2026, preprint, \eprint{2601.00099}.
	
	\bibitem{LitniewskiCiach2021}
	Litniewski~M., Ciach~A., Adsorption in mixtures with competing interactions,
	Molecules, 2021, \textbf{26}, 4532, \doi{10.3390/molecules26154532}.
	
	\bibitem{LitniewskiGozdzCiach2024}
	Litniewski~M., G{\'o}{\'z}d{\'z}~W.~T., Ciach~A., Adsorption on a spherical
	colloidal particle from a mixture of nanoparticles with competing
	interactions, Molecules, 2024, \textbf{29}, 3170,
	\doi{10.3390/molecules29133170}.
	
	\bibitem{LiEtAl2024}
	Li~N., Li~J., Qing~L., Ma~S., Li~Y., Li~B., Self-assembly of colloids with
	competing interactions confined in spheres, Soft Matter, 2024, \textbf{20},
	304, \doi{10.1039/D3SM01227A}.
	
	\bibitem{SernaNoyaGozdz2020}
	Serna~H., Noya~E.~G., G{\'o}{\'z}d{\'z}~W.~T., Confinement of colloids with
	competing interactions in ordered porous materials, J. Phys. Chem. B, 2020,
	\textbf{124}, 10567, \doi{10.1021/acs.jpcb.0c08148}.
	
	\bibitem{Kravtsiv2022}
	Kravtsiv~I., Patsahan~T., Holovko~M., Di~Caprio~D., Soft core fluid with
	competing interactions at a hard wall, J. Mol. Liq., 2022, \textbf{362},
	119652, \doi{10.1016/j.molliq.2022.119652}.
	
	\bibitem{DiCaprio1998}
	Di~Caprio~D., Stafiej~J., Badiali~J.~P., A field theory study of the effect of
	specific interactions in ionic systems: A simple model, J. Chem. Phys., 1998,
	\textbf{108}, 8572, \doi{10.1063/1.476286}.
	
	\bibitem{DiCaprio2001}
	Di~Caprio~D., Stafiej~J., A field theoretical approach to the
	{Debye-H{\"u}ckel} electrolyte differential capacitance in a slab, Condens.
	Matter Phys., 2001, \textbf{4}, 715, \doi{10.5488/CMP.4.4.715}.
	
	\bibitem{DiCaprio2003}
	Di~Caprio~D., Stafiej~J., Badiali~J.~P., Field theoretical approach to
	inhomogeneous ionic systems: thermodynamic consistency with the contact
	theorem, {Gibbs} adsorption and surface tension, Mol. Phys., 2003,
	\textbf{101}, 2545, \doi{10.1080/0026897031000154293}.
	
	\bibitem{DiCaprioBadiali2008}
	Di~Caprio~D., Badiali~J.~P., A formally exact field theory for classical
	systems at equilibrium, J. Phys. A: Math. Theor., 2008, \textbf{41}, 125401,
	\doi{10.1088/1751-8113/41/12/125401}.
	
	\bibitem{HolovkoKravtsivSoviak2009}
	Holovko~M.~F., Kravtsiv~I.~Y., Soviak~E.~M., Spatially confined system
	interacting with {Yukawa} potential, Condens. Matter Phys., 2009,
	\textbf{12}, 137, \doi{10.5488/CMP.12.2.137}.
	
	\bibitem{DiCaprio2011}
	Di~Caprio~D., Stafiej~J., Holovko~M., Kravtsiv~I., {Yukawa} fluid at a hard
	wall: field theory description, Mol. Phys., 2011, \textbf{109}, 695,
	\doi{10.1080/00268976.2010.547524}.
	
	\bibitem{Kravtsiv2015}
	Kravtsiv~I., Patsahan~T., Holovko~M., Di~Caprio~D., Two-{Yukawa} fluid at a
	hard wall: Field theory treatment, J. Chem. Phys., 2015, \textbf{142},
	194708, \doi{10.1063/1.4921242}.
	
	\bibitem{DiCaprio2016}
	Di~Caprio~D., Kravtsiv~I., Patsahan~T., Holovko~M., Oscillating {Yukawa} fluid
	at a hard wall: field theory description, Mol. Phys., 2016, \textbf{114},
	2500, \doi{10.1080/00268976.2016.1174787}.
	
	\bibitem{HolovkoKravtsivDiCaprio2011}
	Holovko~M., Di~Caprio~D., Kravtsiv~I., {Maier-Saupe} nematogenic fluid: field
	theoretical approach, Condens. Matter Phys., 2011, \textbf{14}, 33605,
	\doi{10.5488/CMP.14.33605}.
	
	\bibitem{KravtsivHolovkoDiCaprio2013}
	Kravtsiv~I., Holovko~M., Di~Caprio~D., {Maier--Saupe} nematogenic fluid
	interacting with an isotropic and an anisotropic {Yukawa} potentials: field
	theory description, Mol. Phys., 2013, \textbf{111}, 1023,
	\doi{10.1080/00268976.2012.762615}.
	
	\bibitem{HolovkoKravtsivDiCaprio2013}
	Holovko~M., Kravtsiv~I., Di~Caprio~D., Nematic fluid at a hard wall in the mean
	field approximation, Condens. Matter Phys., 2013, \textbf{16}, 14002,
	\doi{10.5488/CMP.16.14002}.
	
	\bibitem{Holovko2016}
	Holovko~M., Patsahan~T., Kravtsiv~I., Di~Caprio~D., {Maier-Saupe} nematogenic
	fluid with isotropic {Yukawa} repulsion at a hard wall: Mean field
	approximation, Condens. Matter Phys., 2016, \textbf{19}, 13608,
	\doi{10.5488/CMP.19.13608}.
	
	\bibitem{Evans1979}
	Evans~R., The nature of the liquid-vapour interface and other topics in the
	statistical mechanics of non-uniform, classical fluids, Adv. Phys., 1979,
	\textbf{28}, 143, \doi{10.1080/00018737900101365}.
	
	\bibitem{Holovko2023}
	Holovko~M., Vlachy~V., Di~Caprio~D., On the contact conditions for the density
	and charge profiles in the theory of electrical double layer: From planar to
	spherical and cylindrical geometry, J. Mol. Liq., 2023, \textbf{371}, 121040,
	\doi{10.1016/j.molliq.2022.121040}.
	
	\bibitem{allen2017computer}
	Allen~M.~P., Tildesley~D.~J., Computer simulation of liquids, Oxford University
	Press, 2017.
	
	\bibitem{thompson2022lammps}
	Thompson~A.~P., Aktulga~H.~M., Berger~R., Bolintineanu~D.~S., Brown~W.~M.,
	Crozier~P.~S., in't Veld~P.~J., Kohlmeyer~A., Moore~S.~G., Nguyen~T.~D.,
	et~al., {LAMMPS} -- a flexible simulation tool for particle-based materials
	modeling at the atomic, meso, and continuum scales, Comput. Phys. Commun.,
	2022, \textbf{271}, 108171.
	
\end{thebibliography}

\lastpage

\end{document}